\documentclass[
 preprint,
 amsmath,
 amssymb,
 aps, onecolumn,
]{revtex4-1}

\usepackage{graphicx}
\usepackage{dcolumn}
\usepackage{bm}
\usepackage{color, float}

\DeclareMathAlphabet{\pazocal}{OMS}{zplm}{m}{n}

\begin{document}
\title{Quasinormal spectrum of $(2+1)$-dimensional  asymptotically flat, dS and AdS black holes}
\author{Milena Skvortsova}
\email{milenas577@mail.ru}
\affiliation{Peoples' Friendship University of Russia (RUDN University), 6 Miklukho-Maklaya Street, Moscow, 117198, Russia}
\date{\today}
\begin{abstract}
While $(2+1)$-dimensional black holes in the Einstein theory allow for only the anti-de Sitter asymptotic, when the higher curvature correction is tuned on, the asymptotically flat, de Sitter and anti-de Sitter cases are included. Here we propose first comprehensive study of the stability and  quasinormal spectra of the scalar field perturbations around such black holes with all three asymptotics. Calculations of the frequencies are fulfilled with the help of the 6th order WKB method with Pade approximants, Bernstein polynomial method and time-domain integration. Results obtained by all three methods are in a very good agreement in their common range of applicability. When the multipole moment $k$ is equal to zero, the purely imaginary, i.e. non-oscillatory, modes dominate in the spectrum for all types of the asymptotic behavior, while the spectrum at higher $k$ resembles that in four-dimensional spacetime with the corresponding asymptotic.
\end{abstract}


\maketitle
\section{Introduction}\label{intro}

Recent and forthcoming observations of black holes in the gravitational and electromagnetic spectra \cite{LIGOScientific:2016aoc,EventHorizonTelescope:2019dse,Goddi:2016qax,LISACosmologyWorkingGroup:2022jok} make a great impetus to theoretical study of characteristic frequencies of black holes, called {\it quasinormal modes}  \cite{Berti:2009kk,Nollert:1999ji,Kokkotas:1999bd,Konoplya:2011qq}.

While the main object of theoreticians is spectra of $(3+1)$- dimensional black holes, there were always attempts to formulate a simpler, lower dimensional models which would carry the properties of the four-dimensional objects. The most striking example is the $(2+1)$-dimensional asymptotically anti-de Sitter (AdS) black holes suggested in \cite{Banados:1992wn} and called {\it BTZ black holes} after the names of the authors of the metric. Quasinormal modes of of these BTZ spacetimes were extensively studied (see, for instance, \cite{Govindarajan:2000vq,Cardoso:2001hn,Konoplya:2004ik,Fontana:2023dix}, and references therein)  and the spectra proved out to be qualitatively different from the asymptotically flat and de Sitter one. In \cite{Birmingham:2001pj} it was shown that the quasinormal modes of the BTZ black holes \cite{Govindarajan:2000vq,Cardoso:2001hn} exactly coincide with the poles of the retarded Green function in the two-dimensional dual Conformal Field Theory, owing to the AdS/CFT correspondence \cite{Maldacena:1997re}.
Although it was much easier to find the spectrum for the BTZ spacetime (in comparison with the four and higher dimensional solutions), it did not bare the features of the asymptotically flat space and was of limited interest mainly within the realm of holography and quantum gravity of lower-dimensional systems.

The $(2+1)$-dimensional black-hole metric we are interested in this work allows for all three asymptotics: flat, de Sitter and anti-de Sitter.  It was obtained in \cite{Konoplya:2020ibi}  in a similar fashion with the regularization procedure suggested in \cite{Glavan:2019inb} of the Einstein-Gauss-Bonnet equations of motion.  Then it was generalized in \cite{Hennigar:2020fkv,Hennigar:2020drx}. Although it turned out that the naive regularization of \cite{Glavan:2019inb} does not produce a consistent theory, the black hole solution obtained by this way,  satisfied also the well-behaved gravitational theory \cite{Aoki:2020lig}.
The Gauss-Bonnet corrections in the asymptotically AdS black hole spacetimes play an important role when searching for dual hydrodynamic description at intermediate couplings \cite{Grozdanov:2021jfw,Konoplya:2017zwo}.

Here we will study quasinormal modes of the  $(2+1)$-dimensional black-holes \cite{Konoplya:2020ibi} for all three types of asymptotics.  One of the important aspects on which we will concentrate, is dynamical stability of perturbations under consideration. Quasinormal modes were always an effective tool to study (in)stability of black holes (see, for example, \cite{Takahashi:2010gz,Ishihara:2008re,Dyatlov:2010hq}).  In \cite{Skvortsova:2023zca} we have shown that the asymptotically flat perturbative (in the coupling constant $\alpha$) branch is stable, despite the negative gap of the corresponding effective potential. Here we will extend the analysis to the other asymptotics and the non-perturbative branch. We will also study the spectrum of a massive scalar field. The letter is known to possess peculiar features in four dimensions, such as arbitrarily long lived quasinormal modes, called {\it quasiresonances}  which are known to exist for asymptotically flat black holes \cite{Ohashi:2004wr}, but not for asymptotically de Sitter ones \cite{Konoplya:2005et}. Although quasinormal normal modes of massive (or effectively massive) fields have been extensively studied in the background of the four dimensional black holes \cite{Bolokhov:2023bwm,Bolokhov:2023ruj,Konoplya:2007yy,Zinhailo:2018ska,Kokkotas:2010zd}, including even the case with the Gauss-Bonnet correction \cite{Konoplya:2020bxa,Konoplya:2019hml}, no such analysis was performed for the $2+1$ dimensional asymptotically flat or de Sitter black holes.
At the same time, numerious works on quasinormal modes of BTZ-like black holes
\cite{Chen:2023cjd,Panotopoulos:2018can,Huang:2018vlq,Gupta:2017lwk,Prasia:2016esx,Becar:2013qba,Kim:2012pt,Myung:2012sh} did not include higher curvature corrections and were mainly devoted to asymptotically AdS spacetimes.

Here we will compute quasinormal frequencies with the help of three methods: 6th order WKB method with Pade approximants, time-domain integration and the Bernstein polynomial methods. While the WKB method in the form suggested in \cite{wkb1,wkb2,Konoplya:2003ii} cannot be applied to asymptotically AdS spacetimes, the other two methods can be effectively used for all three types of asymptotic behavior. We will show that the results obtained by all the three methods are in a very good agreement in the common range of their applicability.

The paper is organized in the following way. In sec. II we briefly discuss the properties of the black-hole metric under consideration and the wave like equation for a, generally, massive scalar field. Sec. III is devoted the review of methods for calculation of quasinormal frequencies, while the Sec. IV discusses the obtained results for all three asymptotics. Finally, in the Conclusions we summarize the obtained results.

\section{Black hole metrics and wavelike equations}\label{sec.Bardeen spacetime and the wavelike equations}

The metric of the (2+1)-dimensional black hole has the following general line element
\begin{equation}\label{spherical}
\mathrm{d}s^2 = -f(r)\mathrm{d} t^2 + f(r)^{-1} \mathrm{d} r^2 + r^2  \mathrm{d} x ^2,
\end{equation}
where the metric function which includes the Gauss-Bonnet corrections was found in \cite{Konoplya:2020ibi} (see also \cite{Hennigar:2020fkv})
\begin{equation}
f(r) =1-\frac{r^2}{2 \alpha} \left(-1 \pm  \sqrt{\frac{4 \alpha \left(\Lambda
   \left(r^2-r_H^2\right)+\frac{\alpha}{r_H^2}+1\right)}{r^2}+1}\right).
\end{equation}
Here, $r_H$ is the event horizon radius, $\Lambda$ is the cosmological constant, $\alpha$ is the Gauss-Bonnet coupling constant.

The above metric includes two branches: the perturabtive in $\alpha$, once $1+ 2 \alpha/r_{H}^2 >0$ and a "plus" sign before the square root, and the non-perturbative one, if one chooses "minus" and $1+ 2 \alpha/r_{H}^2 <0$.

(a) In the perturbative branch $\Lambda$ is constrained as follows:
\begin{equation}\label{in1}
\Lambda < -\frac{\alpha}{r_H^4},
\end{equation}
which means that if $\alpha \geq 0$ the cosmological constant must be negative and consequently the black hole spacetime must be asymptotically anti-de Sitter.
The asymptotic form of the metric produces the constrain
\begin{equation}\label{in2}
1+ 4 \alpha \Lambda \geq 0.
\end{equation}
At $\alpha < 0$ and $\Lambda =0$, the metric is asymptotically flat, while
if $\alpha \rightarrow 0$, the function $f(r)$ vanishes and the solution does not exist which reflects the fact that at zero coupling constant there are only asymptotically AdS (BTZ \cite{Banados:1992wn} ) black holes. If $\alpha<0$ and $\Lambda >0$, the metric is asymptotically de Sitter and the cosmological constant is constrained according to the above inequality (\ref{in1},\ref{in2}).

(b) In the non-perturbative branch one always has $\alpha <0$, which means the upper limit for the black hole radius,
\begin{equation}
r_H^2 < - 2 \alpha.
\end{equation}
However, the black hole solution appears only if the charge is non-zero and larger than some minimal value determined in \cite{Konoplya:2020ibi}. Here we will consider the neutral case only, so that from here and on we will be limited by the perturbative branch.

\begin{figure}[H]
	\begin{center}
		\begin{tabular}{cc}
			\includegraphics[width=3in]{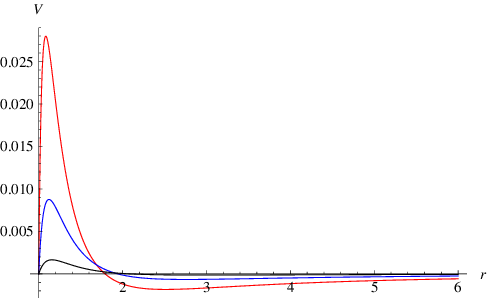}
    \includegraphics[width=3in]{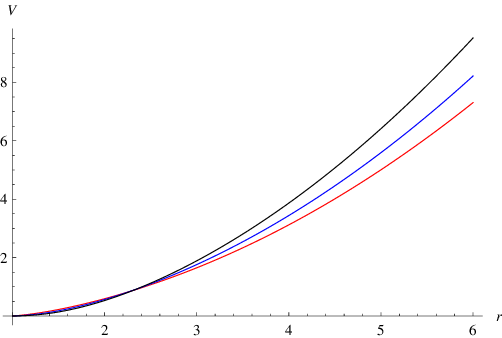}\\
		\end{tabular}
		\caption{Left: Effective potential for asymptotically flat black holes with $k=0$, $\alpha=-0.1$ (bottom, black), $\alpha=-0.2$, and $\alpha=-0.3$ (top, red) $r_H =1$. Right: Effective potential for asymptotically AdS black holes with $k=0$, $\alpha=0.3$ (top, black), $\alpha=0.2$, and $\alpha=0.3$ (bottom, red), $r_H =1$, $\Lambda =-0.5$.  }
		\label{fig:POT}
	\end{center}
\end{figure}

The general-covariant Klein-Gordon equation can be written as follows,
\begin{equation}
\frac{1}{\sqrt{-g}}\partial_\mu \left(\sqrt{-g}g^{\mu \nu}\partial_\nu\Phi\right) =0.
\end{equation}
After the separation of variables it can be reduced to the wave-like equation,
\begin{equation}\label{wave-equation}
\dfrac{d^2 \Psi}{dr_*^2}+(\omega^2-V(r))\Psi=0,
\end{equation}
where the ``tortoise coordinate'' $r_*$ is defined as:
\begin{equation}
dr_*\equiv\frac{dr}{f(r)}.
\end{equation}
The effective potential has the form
\begin{equation}\label{potentialScalar}
V(r)=f(r) \left(\frac{k^2}{r^2}+\frac{f'(r)}{2 r} - \frac{f(r)}{4 r^2} \right),
\end{equation}
where  $k=0, 1, 2,...$ is the multipole number. Examples of the effective potentials for asymptotically flat and AdS black holes at $k=0$ are shown in figs. \ref{fig:POT},\ref{fig:timedomainDS}. The presence of the negative gap for asymptotically flat and de Sitter black holes means that the further study of the stability is necessary for those cases. Higher values of $k$ as well as the asymptotically AdS black hole at all $k$ provide positive definite effective potentials and are therefore guaranteed to be stable.

\section{The three methods for calculation of quasinormal modes}\label{geneqcond}

Quasinormal modes are proper oscillation frequencies with the boundary conditions which are the same at the event horizon, but different, depending on the asymptotic of the spacetime. At the event horizon we always require the purely incoming wave, because classical event horizon (when neglecting the Hawking radiation) has complete absorption of the incoming wave. In far zone the boundary conditions are:
\begin{itemize}
\item purely outgoing wave at infinity or de Sitter horizon for asymptotically flat or de Sitter case respectively, and,
\item vanishing wave function (Dirichlet condition) at the AdS horizon.
\end{itemize}

\subsection{Time-domain integration}

The evolution of perturbations in time can be analyzed via integration of the wave equation at a fixed value of the radial coordinate in time domain. This way contribution of all modes at a given multipole moment is included and the instability, if any, can be detected via observation of growing profiles. Here we will use the time-domain integration method suggested in \cite{Gundlach:1993tp}. This method  was further used in a number of works (see, for example, \cite{Konoplya:2007yy,Churilova:2019cyt,Bronnikov:2019sbx,Ishihara:2008re,Konoplya:2005et}) and showed a very good agreement with the precise methods for the fundamental mode.
The  method for asymptotically flat spacetime is based on the integration of the master wave-like equation in the null-cone coordinates $u=t-r_*$, $v=t+r_*$ and usage of the  discretization scheme \cite{Gundlach:1993tp},
\begin{equation}\label{Discretization}
\Psi\left(N\right)=\Psi\left(W\right)+\Psi\left(E\right)-\Psi\left(S\right) - \Delta^2V\left(S\right)\frac{\Psi\left(W\right)+\Psi\left(E\right)}{4}+{\cal O}\left(\Delta^4\right).
\end{equation}
In the above equation we used the following definition of the points: $N\equiv\left(u+\Delta,v+\Delta\right)$, $W\equiv\left(u+\Delta,v\right)$, $E\equiv\left(u,v+\Delta\right)$, $S\equiv\left(u,v\right)$. Then, the Gaussian wave is supposed to propagate along the null surfaces $u=u_0$ and $v=v_0$. For asymptotically AdS black holes the discretization scheme by Molina et. al. \cite{Wang:2004bv} was used.

\subsection{WKB method}

For checking the results obtained by the time-domain integration at $k>0$ we used the 6th order WKB method with Pade approximants \cite{Konoplya:2019hlu,Matyjasek:2017psv,Konoplya:2003ii}. This method was used in great number of publications (see, for instance, \cite{Bolokhov:2023dxq,Bolokhov:2023ruj,Bolokhov:2023bwm,Konoplya:2019hml}
) showing good concordance with accurate results whenever $k \geq n$ and the effective potential has a single maximum. It is based on the expansion of the wave function in the Taylor series near the peak of the effective potential and further matching it with the asymptotic WKB expansions near the event horizon and at infinity (or de Sitter horizon). This method implies that there are two turning points, so strictly speaking it should not be applied to the massive scalar field, for which the third turning point appears. Nevertheless, comparison with precise calculations shows that usually it provides reasonably good accuracy for relatively small values of the scalar field mass \cite{Konoplya:2019hlu,Konoplya:2007yy}.  The WKB formula of the $n$-th order requires taking of the derivatives of the effective potential up to the $2n$-th order. The explicit form of the WKB formula as well as further details on this method can be found in a review \cite{Konoplya:2019hlu}.

\subsection{Bernstein polynomial method}

For asymptotically flat or AdS black holes, we introduce the compact coordinate $u$, as prescribed in \cite{Fortuna:2020obg},
$$u\equiv\frac{1}{r}.$$
Then, we extract the regular part of the wave function $\Psi(u)$, defined as $y(u)$ multiplied by some function which keeps the singularities, in a similar fashion with \cite{Konoplya:2023aph}, and write $y(u)$ as a sum
\begin{equation}\label{Bernsteinsum}
y(u)=\sum_{k=0}^NC_kB_k^N(u),
\end{equation}
where
$$B_k^N(u)\equiv\frac{N!}{k!(N-k)!}u^k(1-u)^{N-k}$$
are the Bernstein polynomials. For asymtotically de Sitter case the choice of the compact coordinate is different  \cite{Konoplya:2022zav}. Using the Chebyschev collocation grid with $N+1$ points, one finds a set of linear equations with respect to $C_k$, which has nontrivial solutions once the coefficient matrix is singular. Then one has to solve numerically the eigenvalue problem of a matrix pencil with respect to $\omega$ and find coefficients $C_k$. Once  the polynomial (\ref{Bernsteinsum}) is found, it can be used to approximate solutions to the corresponding wave equation \cite{Fortuna:2020obg}. This method is known to be especially effective when searching for the purely imaginary, i.e. non-oscillatory modes. Here we used the package for the Bernstein spectral method which can be applied to all three types of the asymptotic behavior and is publicly available from  \cite{Konoplya:2022zav}.

\section{Quasinormal modes and stability}

Here we will discuss the quasinormal frequencies obtained by all three methods, stability and evolution of perturbations in time domain, including the intermediate and asymptotic tails, for the three types of asymptotic behaviour.

\subsection{Asymptotically flat case}

\begin{figure}[H]
	\begin{center}
		\begin{tabular}{cc}
			\includegraphics[width=3in]{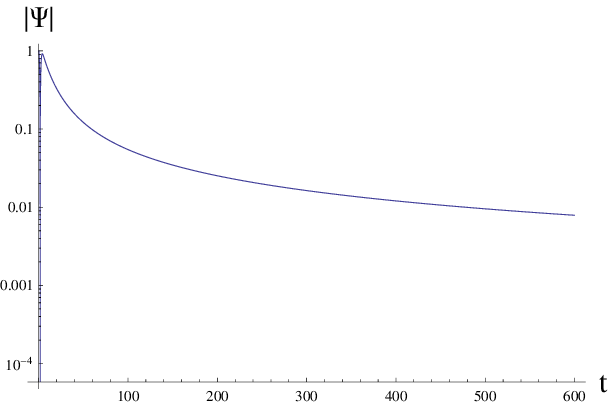}
    \includegraphics[width=3in]{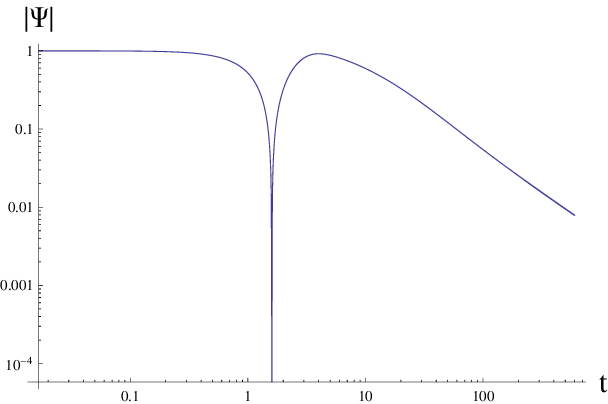}\\
		\end{tabular}
		\caption{Left: Semi-logarithmic plot of absolute value of wave function as a function of time. Right: logarithmic plot. Here we have $k=1$, $\Lambda=0$, $\alpha=-0.49$, $r_H =1$, $\mu =0$.}
		\label{fig:timedomaink0mu0}
	\end{center}
\end{figure}

\begin{figure}[H]
	\begin{center}
		\begin{tabular}{cc}
			\includegraphics[width=3in]{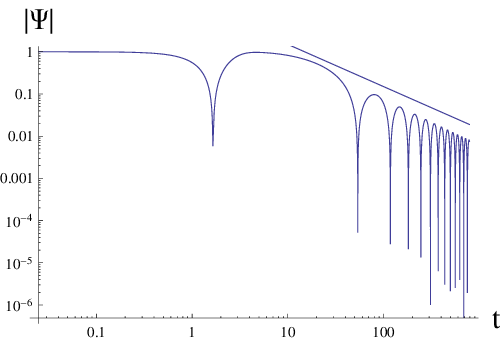}
    \includegraphics[width=3in]{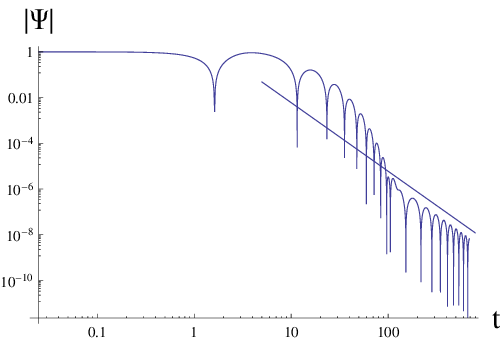}\\
		\end{tabular}
		\caption{Logarithmic plot of absolute value of wave function as a function of time together with the line $\sim t^{-(1+2 k)}$. Here we have $k=0$ (left) and $k=1$ (right), $\Lambda=0$, $\alpha=-0.25$, $r_H =1$, $\mu =0.1$.}
		\label{fig:timedomain12}
	\end{center}
\end{figure}

\begin{figure}[H]
	\begin{center}
		\begin{tabular}{cc}
			\includegraphics[width=3in]{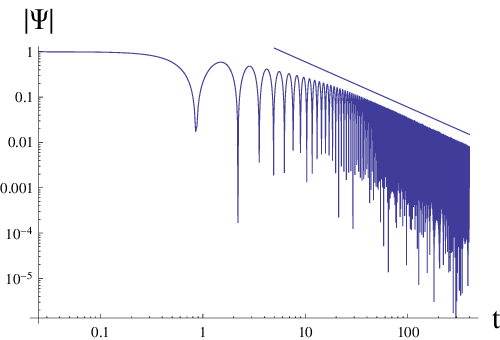}
    \includegraphics[width=3in]{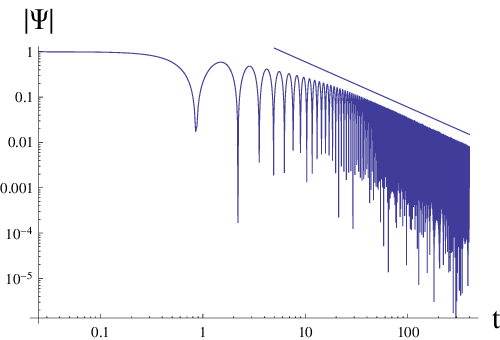}\\
		\end{tabular}
		\caption{Logarithmic plot of absolute value of wave function as a function of time together with the line $\sim t^{-1}$. Here we have $k=0$ (left) and $k=1$ (right), $\Lambda=0$, $\alpha=-0.25$, $r_H =1$, $\mu =5$. At this choice of the parameters, the effective potentials are dominated by the mass term $\mu^2$ and almost indistinguishable. Consequently, the time-domain profiles practically do not differ.}
		\label{fig:timedomain34}
	\end{center}
\end{figure}

\begin{table}
\begin{tabular}{ccccc}
\hline
\hline
 $\alpha$ & WKB-6th  & WKB 6th $(\tilde{m}=4)$ & WKB 6th $(\tilde{m}=5)$ & Time-domain\\
\hline
 -0.05 & \text{0.113474-0.018831 i} &   \text{0.113474-0.018832 i}  & \text{0.113474-0.018832 i} & 0.113475 - 0.0188308 i\\
 -0.1 & \text{0.162438-0.040272 i} & \text{0.162481-0.040287 i}
   & \text{0.162482-0.040284 i} & 0.162494 - 0.0402705 i\\
 -0.15 & \text{0.199705-0.065110 i} & \text{0.200795-0.064978 i}
   & \text{0.200583-0.064855 i} & 0.200706 - 0.0647788 i\\
 -0.2 & \text{0.222710-0.097917 i} & \text{0.232438-0.092848 i}
   & \text{0.231375-0.092840 i} & 0.232197 - 0.09272 i\\
 -0.25 & \text{0.198068-0.179425 i} & \text{0.258296-0.124127 i}
   & \text{0.253298-0.122829 i} & 0.257531 - 0.124026 i\\
 -0.3 & \text{0.147112-0.510272 i} & \text{0.279402-0.159111 i}
   & \text{0.256398-0.143391 i} & 0.276146 - 0.157778 I\\
 -0.35 & \text{0.238216-1.258200 i} & \text{0.294961-0.202321 i}
   & \text{0.182390-0.117272 i} & 0.287704 - 0.191991 i\\
 -0.4 & \text{0.94005-2.17251 i} & \text{0.292964-0.261626 i} &
   \text{0.040493-0.142892 i}& 0.293085 - 0.224448 i\\
 -0.45 & \text{17.3639-0.9183 i} & \text{0.195434-0.333651 i} &
   \text{0.0118962-0.1108563 i}& 0.294249 - 0.253671 i\\ 
 -0.49 & --& -- &--& 0.293476 - 0.274684 i\\ 
\hline
\hline
\end{tabular}
\caption{Asymptotically flat case: Fundamental quasinormal modes ($n=0$) for $k=1$ massless perturbations at $\Lambda =0$ obtained by the WKB method and time-domain integration.}
\end{table}

\begin{table}
\begin{tabular}{cccc}
\hline
\hline
 $\mu$ & \text{6WKB} & WKB 6th ($\tilde{m}=4$) & WKB 6th ($\tilde{m}=5$) \\
\hline
 0. & \text{0.199705-0.065110 i} & \text{0.200795-0.064978 i} &
   \text{0.200583-0.064855 i} \\
 0.05 & \text{0.200163-0.064893 i} & \text{0.201245-0.064626 i}
   & \text{0.201038-0.064638 i} \\
 0.1 & \text{0.201539-0.064246 i} & \text{0.202452-0.063934 i}
   & \text{0.202403-0.063987 i} \\
 0.15 & \text{0.203835-0.063174 i} & \text{0.204691-0.062886 i}
   & \text{0.204680-0.062908 i} \\
 0.2 & \text{0.207053-0.061684 i} & \text{0.207875-0.061398 i}
   & \text{0.207871-0.061409 i} \\
 0.25 & \text{0.211199-0.059790 i} & \text{0.211981-0.059496 i}
   & \text{0.211979-0.059506 i} \\
 0.3 & \text{0.216275-0.057508 i} & \text{0.217008-0.057205 i}
   & \text{0.217007-0.057220 i} \\
 0.35 & \text{0.222288-0.054860 i} & \text{0.222956-0.054548 i}
   & \text{0.222958-0.054575 i} \\
 0.4 & \text{0.229239-0.051872 i} & \text{0.229830-0.051556 i}
   & \text{0.229839-0.051604 i} \\
 0.45 & \text{0.237130-0.048574 i} & \text{0.237630-0.048260 i}
   & \text{0.237656-0.048337 i} \\
 0.5 & \text{0.245964-0.045002 i} & \text{0.246360-0.044698 i}
   & \text{0.246415-0.044809 i} \\
 0.55 & \text{0.255738-0.041196 i} & \text{0.256023-0.040916 i}
   & \text{0.256118-0.041057 i} \\
 0.6 & \text{0.266452-0.037201 i} & \text{0.266627-0.036963 i}
   & \text{0.266765-0.037122 i} \\
 0.65 & \text{0.278104-0.033065 i} & \text{0.278188-0.032889 i}
   & \text{0.278348-0.033052 i} \\
 0.7 & \text{0.290690-0.028841 i} & \text{0.290720-0.028734 i}
   & \text{0.290850-0.028888 i}\\
   \hline
   \hline
\end{tabular}
\caption{Asymptotically flat case: Fundamental quasinormal modes ($n=0$) for $k=1$ massive perturbations obtained by the WKB method; $\alpha =-0.15$.}
\end{table}

Asymptotically flat black holes in the perturbative branch are characterized by two qualitatively different types of modes. The modes corresponding to $k \geq 1$ have both real and imaginary parts representing decaying oscillations, while the $k=0$ modes of the massless field are purely imaginary, i.e. exponentially decaying in time, as can be seen from figs. \ref{fig:timedomaink0mu0}. The $k=1$ modes computed by the WKB and time-domain integration methods are shown in table I. The discrepancy between the methods grows as $\alpha$ approaches its extreme value and as the WKB method converges only asymptotically, but not in each order, the results must be interpreted in favour of the time-domain integration method. In the near extreme regime $\alpha =0.49$ the WKB formula has very large error producing senseless results. Nevertheless, for small and moderate values of the coupling constant the agreement between the two methods is quite good.

 The instability of a static background is governed
 by the non-oscillatory, i.e. purely imaginary, growing mode \cite{Konoplya:2008yy}. Positive definite effective potential guarantee stability of the perturbation. Therefore, $k=1,2,..$ multipoles are stable. Time-domain profiles for $k=0$ shown in \cite{Skvortsova:2023zca}, as well as here in figs. \ref{fig:timedomaink0mu0} signify the stability of perturbations even for the near extreme values of the coupling constant.

When the massive term is turned on, the damping rate is suppressed (see table II) indicating the existence of quasi-resonances, similar to the ones observed in \cite{Ohashi:2004wr} and subsequent papers in four dimensions. Nevertheless, the clear evidence of the arbitrarily long lived modes must be done with the help of a quickly convergent
method of calculation\cite{Zhidenko:2006rs}, such as the Leaver method \cite{Leaver:1985ax}.
Here it cannot be applied to the wave-like equation in its present form, because the Leaver method requires polynomial form of the master differential equation.

In four dimensional spacetimes at asymptotically late times, a massless scalar decay according to the power-law  \cite{Price:1972pw}, while the massive scalar field \cite{Koyama:2000hj,Koyama:2001qw} is oscillatory and the envelope does not depend on the multipole number:
\begin{equation}
|\Psi| \sim t^{-5/6} \sin (\mu t), \quad t \rightarrow \infty, \quad D=4.
\end{equation}
The same law was observed for a number of other backgrounds and spin of the field  \cite{Moderski:2001tk,Konoplya:2006gq,Jing:2004zb,Seahra:2004fg} and others.
At intermediately late times in four dimensions (or, correspondingly, small $\mu M$)  the decay law is
\begin{equation}
|\Psi| \sim t^{-(\frac{8}{6}+\ell)} \sin (A(\mu) t), \quad D=4.
\end{equation}

In $2+1$-dimensional asymptotically flat case we observe that at intermediate late times, as can be seen in figs. \ref{fig:timedomain12}, the decay law is
\begin{equation}
|\Psi| \sim t^{-(1 + 2 k)} \sin (A(\mu) t), \quad D=3.
\end{equation}
The decay law at asymptotic times does not depend on $k$, as can be seen in figs. \ref{fig:timedomain34}:
\begin{equation}
|\Psi| \sim t^{-1} \sin (A(\mu) t), \quad t \rightarrow \infty, \quad D=3.
\end{equation}
The integration until the decayed asymptotic tails demonstrate the stability of the perturbation.
Notice, that the same asymptotic decay law has been recently observed for the Bardeen spacetime \cite{Bolokhov:2023ruj}.

\subsection{Asymptotically de Sitter case}

\begin{figure}[H]
	\begin{center}
		\begin{tabular}{cc}
			\includegraphics[width=3in]{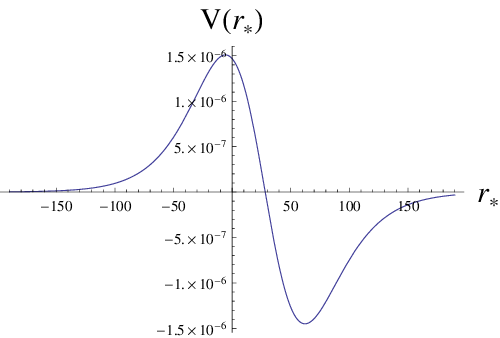}
    \includegraphics[width=3in]{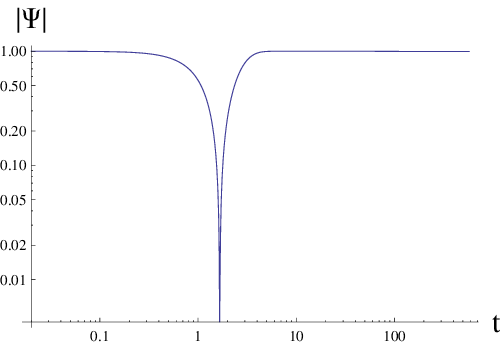}\\
		\end{tabular}
		\caption{Left: Effective potential. Right: Logarithmic plot of absolute value of wave function as a function of time. Here we have $k=0$ (left), $\Lambda=0.24$, $\alpha=-0.25$, $r_H =1$.}
		\label{fig:timedomainDS}
	\end{center}
\end{figure}

In tables III and IV we can see that the positive cosmological constant suppresses both real and imaginary parts of quasinormal modes in the perturbative branch. A similar effect was observed also for asymptotically de Sitter black holes in four and higher dimensions \cite{Zhidenko:2003wq,Konoplya:2004uk,Giammatteo:2005vu}. The time domain profiles shown in figs. \ref{fig:timedomainDS} signifies that the at the near extreme asymptotically de Sitter black holes purely imaginary modes with very small damping rate dominate in the signal. This agrees  with observations in $D=4$  and higher dimensional cases, for which asymptotic tails for Schwarzschild-de Sitter black hole decay exponentially \cite{Molina:2003dc,Brady:1999wd}, which also could be interpreted as dominance of the purely imaginary modes \cite{Konoplya:2022xid}.  However, as in our case the damping rate is tiny in the near extreme regime, there is still a risk of instability in the extreme case.

The massive scalar field does not allow for arbitrarily long-lived quasinormal modes for asymptotically de Sitter black holes. This could be shown analytically exactly in the same way as in \cite{Konoplya:2004wg}.

\begin{table}
\begin{tabular}{ccc}
\hline
\hline
$\Lambda$ & WKB-6th ($\tilde{m}=5$) & Time-domain\\
\hline
0.01 & 0.247756 - 0.125308 i & 0.247676 - 0.123123 i\\
0.05 & 0.203529 - 0.116081 i & 0.206496 - 0.115736 i\\
0.1 &  0.153240 - 0.098081 i & 0.155070 - 0.098524 i\\
0.15 & 0.102442 - 0.073853 i & 0.103707 - 0.074847 i\\
0.2 &  0.051553 - 0.042960 i & 0.051540 - 0.043093 i\\
0.24 & 0.010047 - 0.009668 i & 0.010091 - 0.009704 i\\
\hline
\hline
\end{tabular}
\caption{Asymptotically de Sitter case: Fundamental quasinormal modes ($n=0$) for $k=1$ massless perturbations obtained by the WKB method and time-domain integration; $\alpha =-0.25$.}
\end{table}

\begin{table}
\begin{tabular}{ccc}
\hline
\hline
$\Lambda$ & WKB-6th ($\tilde{m}=5$) & Time-domain\\
\hline
0.01 & 0.543925-0.126157 i & 0.543858 - 0.126648 i\\
0.05 & 0.463678-0.117548 i & 0.463611 - 0.117138 i\\
0.1 &  0.358309-0.099902 i & 0.357864 - 0.101339 i\\
0.15 & 0.246702-0.075731 i & 0.246749 - 0.075648 i\\
0.2 &  0.127664-0.043281 i & 0.127674 - 0.043309 i\\
0.24 & 0.026269-0.009707 i & 0.026269 - 0.009670 i\\
\hline
\hline
\end{tabular}
\caption{Asymptotically de Sitter case: Fundamental quasinormal modes ($n=0$) for $k=2$ massless perturbations obtained by the WKB method and time-domain integration; $\alpha =-0.25$.}
\end{table}

\subsection{Asymptotically anti-de Sitter case}

\begin{figure}[H]
	\begin{center}
		\begin{tabular}{cc}
    \includegraphics[width=3in]{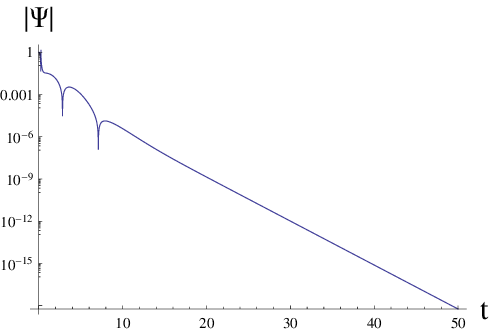}\\
		\end{tabular}
		\caption{Time=domain profile for the asymptotically AdS black hole at $k=1$, $\Lambda=0.1$, $\alpha=-0.5$, $r_H =1$. The dominant mode extracted by the Prony method is $\omega = - 0.719279 i$.}
		\label{fig:timedomainADSk1}
	\end{center}
\end{figure}

This case takes place in the perturbative branch once $\alpha \geq 0$. We apply the time-domain integration and the Bernstein polynomial methods to analyze the evolution of perturbations in this case.

When $\alpha \rightarrow \infty$, the metric function is reduced to the BTZ one \cite{Banados:1992wn},
\begin{equation}
f(r) \rightarrow (r_{H}^{2} - r^2) \Lambda
\end{equation}
and the spectrum known from \cite{Cardoso:2001hn} is reproduced with great accuracy by the Bernstein polynomial method, as can be seen, for example, from table V. The general exact analytical formula for the quasinormal modes of the BTZ black holes in the units we are using here has the following form:
\begin{equation}
\omega_n = \pm \sqrt{\Lambda} k - 2 \Lambda (n+1) i,
\end{equation}
which coincides with  \cite{Cardoso:2001hn}, if one goes over to the units of the AdS radius $l =1/\sqrt{\Lambda}$ and uses $M=r_{H}^2/l^2$.

\begin{table}
\begin{tabular}{ccccc}
\hline
\hline
$\Lambda$ & $n=0$ (B) & $n=1$ (B) & $n=2$ (B) & $n=0$ (T-D) \\
\hline
-0.5 & 0.707107 - 0.999999 i &  0.707107 - 2.000000 i & 0.707107 - 3.000000 i & 0.707130 - 0.999985 i\\
-1 & 1.000000  - 2.000000 i & 0.999999 - 4.000000  i & 1.000000 - 5.999999 i &  1.000225 - 1.999804 i\\
-1.5 & 1.224745 - 3.000000 i &  1.224745 -6.000000 i & 1.224746 -9.000000  i &  1.225700 - 2.994060 i\\
-2 &   1.414214 - 4.000000  i & 1.414214 -8.000000 i & 1.414220 - 12.00000 i &  1.409750 - 3.996330 i\\
\hline
\hline
\end{tabular}
\caption{Asymptotically anti-de Sitter (BTZ) black hole: Quasinormal modes for $k=1$ massless perturbations obtained by the Bernstein polynomial method (B) and time-domain integration (T-D); $\alpha =0$, $r_H=1$.}
\end{table}

\begin{table}
\begin{tabular}{cccc}
\hline
\hline
$\Lambda$ & $\alpha =0.1$ & $\alpha =0.2$ & $\alpha =0.3$\\
\hline
-0.5 &  - 0.719279 i&  - 0.387343 i & - 0.212103 i\\ 
-1 &  - 1.351860  i&  - 0.902496 I & --\\ 
-1.5 &   - 1.87552 i & --& --\\
-2 &  - 2.36785  i& -- & --\\ 
\hline
\hline
\end{tabular}
\caption{Asymptotically anti-de Sitter case: Quasinormal modes for $k=1$ massless perturbations obtained by the time-domain integration. Empty space corresponds to the regime $1+ 4 \alpha \Lambda < 0$ which is outside the constrain given by eq. (\ref{in2}).}
\end{table}

From table VI one can see that the fundamental modes are purely imaginary, which agrees with the exponentially decaying time-domain profile in fig. \ref{fig:timedomainADSk1},  and roughly proportional to $\Lambda$.

\section{Conclusions}\label{conclus}

Here we have analyzed quasinormal modes of a scalar field in the background of the asymptotically flat, de Sitter and anti-de Sitter black holes inspired by the Gauss-Bonnet correction to the Einstein action. Usually $2+1$-dimensional black holes allow for only anti-de Sitter asymptotic behavior. We have shown that the quasinormal spectrum of the $2+1$ asymptotically flat and de Sitter black holes keeps a number of features of the correspondingly asymptotically flat or de Sitter four dimensional spacetimes. In particular,

\begin{itemize}
\item Perturbations of asymptotically flat black holes is not equidistant in $n$, like it happens for the asymptotically AdS case, but resembles the spectrum of the four and higher dimensional asymptotically flat black holes. The same is true for asymptotically de Sitter case.
\item At late times, when the quasinormal modes for asymptotically flat case are suppressed by the late-time tails, the field decays at $k>0$ according to the power law, though via a different from the four-dimensional Price's law \cite{Price:1972pw}.
Nevertheless, $k=0$ perturbations are exceptional, as they are governed by the purely imaginary, i.e. non-oscillatory, exponentially decaying modes at all times.
\item When the massive term is tuned on, an arbitrarily long-lived modes exist for asymptotically flat case, which are called quasi-resonances and extensively studied in four dimensions.
\item The positive cosmological constant suppresses the real oscillation frequency and the damping rate of the perturbation.
\end{itemize}

Despite the deep negative gap of the $k=0$ effective potential, the time-domain integration did not show any instability, but instead showed stable modes with tiny decay rate, which means that further study (possibly with alternative more accurate and stable methods) of the stability in the near extreme regime might be necessary.

\acknowledgments{
The author acknowledges R. A. Konoplya for fruitful discussions, and for careful reading of the manuscript. This work was supported by RUDN University research project FSSF-2023-0003.}

\bibliographystyle{apsrev4-1}
\bibliography{ragsamp}

\begin{thebibliography}{73}%
\makeatletter
\providecommand \@ifxundefined [1]{%
 \@ifx{#1\undefined}
}%
\providecommand \@ifnum [1]{%
 \ifnum #1\expandafter \@firstoftwo
 \else \expandafter \@secondoftwo
 \fi
}%
\providecommand \@ifx [1]{%
 \ifx #1\expandafter \@firstoftwo
 \else \expandafter \@secondoftwo
 \fi
}%
\providecommand \natexlab [1]{#1}%
\providecommand \enquote  [1]{``#1''}%
\providecommand \bibnamefont  [1]{#1}%
\providecommand \bibfnamefont [1]{#1}%
\providecommand \citenamefont [1]{#1}%
\providecommand \href@noop [0]{\@secondoftwo}%
\providecommand \href [0]{\begingroup \@sanitize@url \@href}%
\providecommand \@href[1]{\@@startlink{#1}\@@href}%
\providecommand \@@href[1]{\endgroup#1\@@endlink}%
\providecommand \@sanitize@url [0]{\catcode `\\12\catcode `\$12\catcode
  `\&12\catcode `\#12\catcode `\^12\catcode `\_12\catcode `\%12\relax}%
\providecommand \@@startlink[1]{}%
\providecommand \@@endlink[0]{}%
\providecommand \url  [0]{\begingroup\@sanitize@url \@url }%
\providecommand \@url [1]{\endgroup\@href {#1}{\urlprefix }}%
\providecommand \urlprefix  [0]{URL }%
\providecommand \Eprint [0]{\href }%
\providecommand \doibase [0]{http://dx.doi.org/}%
\providecommand \selectlanguage [0]{\@gobble}%
\providecommand \bibinfo  [0]{\@secondoftwo}%
\providecommand \bibfield  [0]{\@secondoftwo}%
\providecommand \translation [1]{[#1]}%
\providecommand \BibitemOpen [0]{}%
\providecommand \bibitemStop [0]{}%
\providecommand \bibitemNoStop [0]{.\EOS\space}%
\providecommand \EOS [0]{\spacefactor3000\relax}%
\providecommand \BibitemShut  [1]{\csname bibitem#1\endcsname}%
\let\auto@bib@innerbib\@empty
\bibitem [{\citenamefont {Abbott}\ \emph {et~al.}(2016)\citenamefont {Abbott}
  \emph {et~al.}}]{LIGOScientific:2016aoc}%
  \BibitemOpen
  \bibfield  {author} {\bibinfo {author} {\bibfnamefont {B.~P.}\ \bibnamefont
  {Abbott}} \emph {et~al.} (\bibinfo {collaboration} {LIGO Scientific,
  Virgo}),\ }\href {\doibase 10.1103/PhysRevLett.116.061102} {\bibfield
  {journal} {\bibinfo  {journal} {Phys. Rev. Lett.}\ }\textbf {\bibinfo
  {volume} {116}},\ \bibinfo {pages} {061102} (\bibinfo {year} {2016})},\
  \Eprint {http://arxiv.org/abs/1602.03837} {arXiv:1602.03837 [gr-qc]}
  \BibitemShut {NoStop}%
\bibitem [{\citenamefont {Akiyama}\ \emph {et~al.}(2019)\citenamefont {Akiyama}
  \emph {et~al.}}]{EventHorizonTelescope:2019dse}%
  \BibitemOpen
  \bibfield  {author} {\bibinfo {author} {\bibfnamefont {K.}~\bibnamefont
  {Akiyama}} \emph {et~al.} (\bibinfo {collaboration} {Event Horizon
  Telescope}),\ }\href {\doibase 10.3847/2041-8213/ab0ec7} {\bibfield
  {journal} {\bibinfo  {journal} {Astrophys. J. Lett.}\ }\textbf {\bibinfo
  {volume} {875}},\ \bibinfo {pages} {L1} (\bibinfo {year} {2019})},\ \Eprint
  {http://arxiv.org/abs/1906.11238} {arXiv:1906.11238 [astro-ph.GA]}
  \BibitemShut {NoStop}%
\bibitem [{\citenamefont {Goddi}\ \emph {et~al.}(2016)\citenamefont {Goddi}
  \emph {et~al.}}]{Goddi:2016qax}%
  \BibitemOpen
  \bibfield  {author} {\bibinfo {author} {\bibfnamefont {C.}~\bibnamefont
  {Goddi}} \emph {et~al.},\ }\href {\doibase 10.1142/9789813226609_0046}
  {\bibfield  {journal} {\bibinfo  {journal} {Int. J. Mod. Phys. D}\ }\textbf
  {\bibinfo {volume} {26}},\ \bibinfo {pages} {1730001} (\bibinfo {year}
  {2016})},\ \Eprint {http://arxiv.org/abs/1606.08879} {arXiv:1606.08879
  [astro-ph.HE]} \BibitemShut {NoStop}%
\bibitem [{\citenamefont {Auclair}\ \emph {et~al.}(2023)\citenamefont {Auclair}
  \emph {et~al.}}]{LISACosmologyWorkingGroup:2022jok}%
  \BibitemOpen
  \bibfield  {author} {\bibinfo {author} {\bibfnamefont {P.}~\bibnamefont
  {Auclair}} \emph {et~al.} (\bibinfo {collaboration} {LISA Cosmology Working
  Group}),\ }\href {\doibase 10.1007/s41114-023-00045-2} {\bibfield  {journal}
  {\bibinfo  {journal} {Living Rev. Rel.}\ }\textbf {\bibinfo {volume} {26}},\
  \bibinfo {pages} {5} (\bibinfo {year} {2023})},\ \Eprint
  {http://arxiv.org/abs/2204.05434} {arXiv:2204.05434 [astro-ph.CO]}
  \BibitemShut {NoStop}%
\bibitem [{\citenamefont {Berti}\ \emph {et~al.}(2009)\citenamefont {Berti},
  \citenamefont {Cardoso},\ and\ \citenamefont {Starinets}}]{Berti:2009kk}%
  \BibitemOpen
  \bibfield  {author} {\bibinfo {author} {\bibfnamefont {E.}~\bibnamefont
  {Berti}}, \bibinfo {author} {\bibfnamefont {V.}~\bibnamefont {Cardoso}}, \
  and\ \bibinfo {author} {\bibfnamefont {A.~O.}\ \bibnamefont {Starinets}},\
  }\href {\doibase 10.1088/0264-9381/26/16/163001} {\bibfield  {journal}
  {\bibinfo  {journal} {Class. Quant. Grav.}\ }\textbf {\bibinfo {volume}
  {26}},\ \bibinfo {pages} {163001} (\bibinfo {year} {2009})},\ \Eprint
  {http://arxiv.org/abs/0905.2975} {arXiv:0905.2975 [gr-qc]} \BibitemShut
  {NoStop}%
\bibitem [{\citenamefont {Nollert}(1999)}]{Nollert:1999ji}%
  \BibitemOpen
  \bibfield  {author} {\bibinfo {author} {\bibfnamefont {H.-P.}\ \bibnamefont
  {Nollert}},\ }\href {\doibase 10.1088/0264-9381/16/12/201} {\bibfield
  {journal} {\bibinfo  {journal} {Class. Quant. Grav.}\ }\textbf {\bibinfo
  {volume} {16}},\ \bibinfo {pages} {R159} (\bibinfo {year}
  {1999})}\BibitemShut {NoStop}%
\bibitem [{\citenamefont {Kokkotas}\ and\ \citenamefont
  {Schmidt}(1999)}]{Kokkotas:1999bd}%
  \BibitemOpen
  \bibfield  {author} {\bibinfo {author} {\bibfnamefont {K.~D.}\ \bibnamefont
  {Kokkotas}}\ and\ \bibinfo {author} {\bibfnamefont {B.~G.}\ \bibnamefont
  {Schmidt}},\ }\href {\doibase 10.12942/lrr-1999-2} {\bibfield  {journal}
  {\bibinfo  {journal} {Living Rev. Rel.}\ }\textbf {\bibinfo {volume} {2}},\
  \bibinfo {pages} {2} (\bibinfo {year} {1999})},\ \Eprint
  {http://arxiv.org/abs/gr-qc/9909058} {arXiv:gr-qc/9909058} \BibitemShut
  {NoStop}%
\bibitem [{\citenamefont {Konoplya}\ and\ \citenamefont
  {Zhidenko}(2011)}]{Konoplya:2011qq}%
  \BibitemOpen
  \bibfield  {author} {\bibinfo {author} {\bibfnamefont {R.~A.}\ \bibnamefont
  {Konoplya}}\ and\ \bibinfo {author} {\bibfnamefont {A.}~\bibnamefont
  {Zhidenko}},\ }\href {\doibase 10.1103/RevModPhys.83.793} {\bibfield
  {journal} {\bibinfo  {journal} {Rev. Mod. Phys.}\ }\textbf {\bibinfo {volume}
  {83}},\ \bibinfo {pages} {793} (\bibinfo {year} {2011})},\ \Eprint
  {http://arxiv.org/abs/1102.4014} {arXiv:1102.4014 [gr-qc]} \BibitemShut
  {NoStop}%
\bibitem [{\citenamefont {Banados}\ \emph {et~al.}(1992)\citenamefont
  {Banados}, \citenamefont {Teitelboim},\ and\ \citenamefont
  {Zanelli}}]{Banados:1992wn}%
  \BibitemOpen
  \bibfield  {author} {\bibinfo {author} {\bibfnamefont {M.}~\bibnamefont
  {Banados}}, \bibinfo {author} {\bibfnamefont {C.}~\bibnamefont {Teitelboim}},
  \ and\ \bibinfo {author} {\bibfnamefont {J.}~\bibnamefont {Zanelli}},\ }\href
  {\doibase 10.1103/PhysRevLett.69.1849} {\bibfield  {journal} {\bibinfo
  {journal} {Phys. Rev. Lett.}\ }\textbf {\bibinfo {volume} {69}},\ \bibinfo
  {pages} {1849} (\bibinfo {year} {1992})},\ \Eprint
  {http://arxiv.org/abs/hep-th/9204099} {arXiv:hep-th/9204099} \BibitemShut
  {NoStop}%
\bibitem [{\citenamefont {Govindarajan}\ and\ \citenamefont
  {Suneeta}(2001)}]{Govindarajan:2000vq}%
  \BibitemOpen
  \bibfield  {author} {\bibinfo {author} {\bibfnamefont {T.~R.}\ \bibnamefont
  {Govindarajan}}\ and\ \bibinfo {author} {\bibfnamefont {V.}~\bibnamefont
  {Suneeta}},\ }\href {\doibase 10.1088/0264-9381/18/2/306} {\bibfield
  {journal} {\bibinfo  {journal} {Class. Quant. Grav.}\ }\textbf {\bibinfo
  {volume} {18}},\ \bibinfo {pages} {265} (\bibinfo {year} {2001})},\ \Eprint
  {http://arxiv.org/abs/gr-qc/0007084} {arXiv:gr-qc/0007084} \BibitemShut
  {NoStop}%
\bibitem [{\citenamefont {Cardoso}\ and\ \citenamefont
  {Lemos}(2001)}]{Cardoso:2001hn}%
  \BibitemOpen
  \bibfield  {author} {\bibinfo {author} {\bibfnamefont {V.}~\bibnamefont
  {Cardoso}}\ and\ \bibinfo {author} {\bibfnamefont {J.~P.~S.}\ \bibnamefont
  {Lemos}},\ }\href {\doibase 10.1103/PhysRevD.63.124015} {\bibfield  {journal}
  {\bibinfo  {journal} {Phys. Rev. D}\ }\textbf {\bibinfo {volume} {63}},\
  \bibinfo {pages} {124015} (\bibinfo {year} {2001})},\ \Eprint
  {http://arxiv.org/abs/gr-qc/0101052} {arXiv:gr-qc/0101052} \BibitemShut
  {NoStop}%
\bibitem [{\citenamefont {Konoplya}(2004)}]{Konoplya:2004ik}%
  \BibitemOpen
  \bibfield  {author} {\bibinfo {author} {\bibfnamefont {R.~A.}\ \bibnamefont
  {Konoplya}},\ }\href {\doibase 10.1103/PhysRevD.70.047503} {\bibfield
  {journal} {\bibinfo  {journal} {Phys. Rev. D}\ }\textbf {\bibinfo {volume}
  {70}},\ \bibinfo {pages} {047503} (\bibinfo {year} {2004})},\ \Eprint
  {http://arxiv.org/abs/hep-th/0406100} {arXiv:hep-th/0406100} \BibitemShut
  {NoStop}%
\bibitem [{\citenamefont {Fontana}(2023)}]{Fontana:2023dix}%
  \BibitemOpen
  \bibfield  {author} {\bibinfo {author} {\bibfnamefont {R.~D.~B.}\
  \bibnamefont {Fontana}},\ }\href@noop {} {\  (\bibinfo {year} {2023})},\
  \Eprint {http://arxiv.org/abs/2306.02504} {arXiv:2306.02504 [gr-qc]}
  \BibitemShut {NoStop}%
\bibitem [{\citenamefont {Birmingham}\ \emph {et~al.}(2002)\citenamefont
  {Birmingham}, \citenamefont {Sachs},\ and\ \citenamefont
  {Solodukhin}}]{Birmingham:2001pj}%
  \BibitemOpen
  \bibfield  {author} {\bibinfo {author} {\bibfnamefont {D.}~\bibnamefont
  {Birmingham}}, \bibinfo {author} {\bibfnamefont {I.}~\bibnamefont {Sachs}}, \
  and\ \bibinfo {author} {\bibfnamefont {S.~N.}\ \bibnamefont {Solodukhin}},\
  }\href {\doibase 10.1103/PhysRevLett.88.151301} {\bibfield  {journal}
  {\bibinfo  {journal} {Phys. Rev. Lett.}\ }\textbf {\bibinfo {volume} {88}},\
  \bibinfo {pages} {151301} (\bibinfo {year} {2002})},\ \Eprint
  {http://arxiv.org/abs/hep-th/0112055} {arXiv:hep-th/0112055} \BibitemShut
  {NoStop}%
\bibitem [{\citenamefont {Maldacena}(1998)}]{Maldacena:1997re}%
  \BibitemOpen
  \bibfield  {author} {\bibinfo {author} {\bibfnamefont {J.~M.}\ \bibnamefont
  {Maldacena}},\ }\href {\doibase 10.4310/ATMP.1998.v2.n2.a1} {\bibfield
  {journal} {\bibinfo  {journal} {Adv. Theor. Math. Phys.}\ }\textbf {\bibinfo
  {volume} {2}},\ \bibinfo {pages} {231} (\bibinfo {year} {1998})},\ \Eprint
  {http://arxiv.org/abs/hep-th/9711200} {arXiv:hep-th/9711200} \BibitemShut
  {NoStop}%
\bibitem [{\citenamefont {Konoplya}\ and\ \citenamefont
  {Zhidenko}(2020)}]{Konoplya:2020ibi}%
  \BibitemOpen
  \bibfield  {author} {\bibinfo {author} {\bibfnamefont {R.~A.}\ \bibnamefont
  {Konoplya}}\ and\ \bibinfo {author} {\bibfnamefont {A.}~\bibnamefont
  {Zhidenko}},\ }\href {\doibase 10.1103/PhysRevD.102.064004} {\bibfield
  {journal} {\bibinfo  {journal} {Phys. Rev. D}\ }\textbf {\bibinfo {volume}
  {102}},\ \bibinfo {pages} {064004} (\bibinfo {year} {2020})},\ \Eprint
  {http://arxiv.org/abs/2003.12171} {arXiv:2003.12171 [gr-qc]} \BibitemShut
  {NoStop}%
\bibitem [{\citenamefont {Glavan}\ and\ \citenamefont
  {Lin}(2020)}]{Glavan:2019inb}%
  \BibitemOpen
  \bibfield  {author} {\bibinfo {author} {\bibfnamefont {D.}~\bibnamefont
  {Glavan}}\ and\ \bibinfo {author} {\bibfnamefont {C.}~\bibnamefont {Lin}},\
  }\href {\doibase 10.1103/PhysRevLett.124.081301} {\bibfield  {journal}
  {\bibinfo  {journal} {Phys. Rev. Lett.}\ }\textbf {\bibinfo {volume} {124}},\
  \bibinfo {pages} {081301} (\bibinfo {year} {2020})},\ \Eprint
  {http://arxiv.org/abs/1905.03601} {arXiv:1905.03601 [gr-qc]} \BibitemShut
  {NoStop}%
\bibitem [{\citenamefont {Hennigar}\ \emph {et~al.}(2020)\citenamefont
  {Hennigar}, \citenamefont {Kubiznak}, \citenamefont {Mann},\ and\
  \citenamefont {Pollack}}]{Hennigar:2020fkv}%
  \BibitemOpen
  \bibfield  {author} {\bibinfo {author} {\bibfnamefont {R.~A.}\ \bibnamefont
  {Hennigar}}, \bibinfo {author} {\bibfnamefont {D.}~\bibnamefont {Kubiznak}},
  \bibinfo {author} {\bibfnamefont {R.~B.}\ \bibnamefont {Mann}}, \ and\
  \bibinfo {author} {\bibfnamefont {C.}~\bibnamefont {Pollack}},\ }\href
  {\doibase 10.1016/j.physletb.2020.135657} {\bibfield  {journal} {\bibinfo
  {journal} {Phys. Lett. B}\ }\textbf {\bibinfo {volume} {808}},\ \bibinfo
  {pages} {135657} (\bibinfo {year} {2020})},\ \Eprint
  {http://arxiv.org/abs/2004.12995} {arXiv:2004.12995 [gr-qc]} \BibitemShut
  {NoStop}%
\bibitem [{\citenamefont {Hennigar}\ \emph {et~al.}(2021)\citenamefont
  {Hennigar}, \citenamefont {Kubiznak},\ and\ \citenamefont
  {Mann}}]{Hennigar:2020drx}%
  \BibitemOpen
  \bibfield  {author} {\bibinfo {author} {\bibfnamefont {R.~A.}\ \bibnamefont
  {Hennigar}}, \bibinfo {author} {\bibfnamefont {D.}~\bibnamefont {Kubiznak}},
  \ and\ \bibinfo {author} {\bibfnamefont {R.~B.}\ \bibnamefont {Mann}},\
  }\href {\doibase 10.1088/1361-6382/abce48} {\bibfield  {journal} {\bibinfo
  {journal} {Class. Quant. Grav.}\ }\textbf {\bibinfo {volume} {38}},\ \bibinfo
  {pages} {03LT01} (\bibinfo {year} {2021})},\ \Eprint
  {http://arxiv.org/abs/2005.13732} {arXiv:2005.13732 [gr-qc]} \BibitemShut
  {NoStop}%
\bibitem [{\citenamefont {Aoki}\ \emph {et~al.}(2020)\citenamefont {Aoki},
  \citenamefont {Gorji},\ and\ \citenamefont {Mukohyama}}]{Aoki:2020lig}%
  \BibitemOpen
  \bibfield  {author} {\bibinfo {author} {\bibfnamefont {K.}~\bibnamefont
  {Aoki}}, \bibinfo {author} {\bibfnamefont {M.~A.}\ \bibnamefont {Gorji}}, \
  and\ \bibinfo {author} {\bibfnamefont {S.}~\bibnamefont {Mukohyama}},\ }\href
  {\doibase 10.1016/j.physletb.2020.135843} {\bibfield  {journal} {\bibinfo
  {journal} {Phys. Lett. B}\ }\textbf {\bibinfo {volume} {810}},\ \bibinfo
  {pages} {135843} (\bibinfo {year} {2020})},\ \Eprint
  {http://arxiv.org/abs/2005.03859} {arXiv:2005.03859 [gr-qc]} \BibitemShut
  {NoStop}%
\bibitem [{\citenamefont {Grozdanov}\ \emph {et~al.}(2021)\citenamefont
  {Grozdanov}, \citenamefont {Starinets},\ and\ \citenamefont
  {Tadi\'c}}]{Grozdanov:2021jfw}%
  \BibitemOpen
  \bibfield  {author} {\bibinfo {author} {\bibfnamefont {S.}~\bibnamefont
  {Grozdanov}}, \bibinfo {author} {\bibfnamefont {A.~O.}\ \bibnamefont
  {Starinets}}, \ and\ \bibinfo {author} {\bibfnamefont {P.}~\bibnamefont
  {Tadi\'c}},\ }\href {\doibase 10.1007/JHEP06(2021)180} {\bibfield  {journal}
  {\bibinfo  {journal} {JHEP}\ }\textbf {\bibinfo {volume} {06}},\ \bibinfo
  {pages} {180} (\bibinfo {year} {2021})},\ \Eprint
  {http://arxiv.org/abs/2104.11035} {arXiv:2104.11035 [hep-th]} \BibitemShut
  {NoStop}%
\bibitem [{\citenamefont {Konoplya}\ and\ \citenamefont
  {Zhidenko}(2017)}]{Konoplya:2017zwo}%
  \BibitemOpen
  \bibfield  {author} {\bibinfo {author} {\bibfnamefont {R.~A.}\ \bibnamefont
  {Konoplya}}\ and\ \bibinfo {author} {\bibfnamefont {A.}~\bibnamefont
  {Zhidenko}},\ }\href {\doibase 10.1007/JHEP09(2017)139} {\bibfield  {journal}
  {\bibinfo  {journal} {JHEP}\ }\textbf {\bibinfo {volume} {09}},\ \bibinfo
  {pages} {139} (\bibinfo {year} {2017})},\ \Eprint
  {http://arxiv.org/abs/1705.07732} {arXiv:1705.07732 [hep-th]} \BibitemShut
  {NoStop}%
\bibitem [{\citenamefont {Takahashi}\ and\ \citenamefont
  {Soda}(2010)}]{Takahashi:2010gz}%
  \BibitemOpen
  \bibfield  {author} {\bibinfo {author} {\bibfnamefont {T.}~\bibnamefont
  {Takahashi}}\ and\ \bibinfo {author} {\bibfnamefont {J.}~\bibnamefont
  {Soda}},\ }\href {\doibase 10.1143/PTP.124.711} {\bibfield  {journal}
  {\bibinfo  {journal} {Prog. Theor. Phys.}\ }\textbf {\bibinfo {volume}
  {124}},\ \bibinfo {pages} {711} (\bibinfo {year} {2010})},\ \Eprint
  {http://arxiv.org/abs/1008.1618} {arXiv:1008.1618 [gr-qc]} \BibitemShut
  {NoStop}%
\bibitem [{\citenamefont {Ishihara}\ \emph {et~al.}(2008)\citenamefont
  {Ishihara}, \citenamefont {Kimura}, \citenamefont {Konoplya}, \citenamefont
  {Murata}, \citenamefont {Soda},\ and\ \citenamefont
  {Zhidenko}}]{Ishihara:2008re}%
  \BibitemOpen
  \bibfield  {author} {\bibinfo {author} {\bibfnamefont {H.}~\bibnamefont
  {Ishihara}}, \bibinfo {author} {\bibfnamefont {M.}~\bibnamefont {Kimura}},
  \bibinfo {author} {\bibfnamefont {R.~A.}\ \bibnamefont {Konoplya}}, \bibinfo
  {author} {\bibfnamefont {K.}~\bibnamefont {Murata}}, \bibinfo {author}
  {\bibfnamefont {J.}~\bibnamefont {Soda}}, \ and\ \bibinfo {author}
  {\bibfnamefont {A.}~\bibnamefont {Zhidenko}},\ }\href {\doibase
  10.1103/PhysRevD.77.084019} {\bibfield  {journal} {\bibinfo  {journal} {Phys.
  Rev. D}\ }\textbf {\bibinfo {volume} {77}},\ \bibinfo {pages} {084019}
  (\bibinfo {year} {2008})},\ \Eprint {http://arxiv.org/abs/0802.0655}
  {arXiv:0802.0655 [hep-th]} \BibitemShut {NoStop}%
\bibitem [{\citenamefont {Dyatlov}(2011)}]{Dyatlov:2010hq}%
  \BibitemOpen
  \bibfield  {author} {\bibinfo {author} {\bibfnamefont {S.}~\bibnamefont
  {Dyatlov}},\ }\href {\doibase 10.1007/s00220-011-1286-x} {\bibfield
  {journal} {\bibinfo  {journal} {Commun. Math. Phys.}\ }\textbf {\bibinfo
  {volume} {306}},\ \bibinfo {pages} {119} (\bibinfo {year} {2011})},\ \Eprint
  {http://arxiv.org/abs/1003.6128} {arXiv:1003.6128 [math.AP]} \BibitemShut
  {NoStop}%
\bibitem [{\citenamefont {Skvortsova}(2023)}]{Skvortsova:2023zca}%
  \BibitemOpen
  \bibfield  {author} {\bibinfo {author} {\bibfnamefont {M.}~\bibnamefont
  {Skvortsova}},\ }\href@noop {} {\  (\bibinfo {year} {2023})},\ \Eprint
  {http://arxiv.org/abs/2311.02729} {arXiv:2311.02729 [gr-qc]} \BibitemShut
  {NoStop}%
\bibitem [{\citenamefont {Ohashi}\ and\ \citenamefont
  {Sakagami}(2004)}]{Ohashi:2004wr}%
  \BibitemOpen
  \bibfield  {author} {\bibinfo {author} {\bibfnamefont {A.}~\bibnamefont
  {Ohashi}}\ and\ \bibinfo {author} {\bibfnamefont {M.-a.}\ \bibnamefont
  {Sakagami}},\ }\href {\doibase 10.1088/0264-9381/21/16/010} {\bibfield
  {journal} {\bibinfo  {journal} {Class. Quant. Grav.}\ }\textbf {\bibinfo
  {volume} {21}},\ \bibinfo {pages} {3973} (\bibinfo {year} {2004})},\ \Eprint
  {http://arxiv.org/abs/gr-qc/0407009} {arXiv:gr-qc/0407009} \BibitemShut
  {NoStop}%
\bibitem [{\citenamefont {Konoplya}\ and\ \citenamefont
  {Molina}(2005)}]{Konoplya:2005et}%
  \BibitemOpen
  \bibfield  {author} {\bibinfo {author} {\bibfnamefont {R.~A.}\ \bibnamefont
  {Konoplya}}\ and\ \bibinfo {author} {\bibfnamefont {C.}~\bibnamefont
  {Molina}},\ }\href {\doibase 10.1103/PhysRevD.71.124009} {\bibfield
  {journal} {\bibinfo  {journal} {Phys. Rev. D}\ }\textbf {\bibinfo {volume}
  {71}},\ \bibinfo {pages} {124009} (\bibinfo {year} {2005})},\ \Eprint
  {http://arxiv.org/abs/gr-qc/0504139} {arXiv:gr-qc/0504139} \BibitemShut
  {NoStop}%
\bibitem [{\citenamefont {Bolokhov}(2023{\natexlab{a}})}]{Bolokhov:2023bwm}%
  \BibitemOpen
  \bibfield  {author} {\bibinfo {author} {\bibfnamefont {S.~V.}\ \bibnamefont
  {Bolokhov}},\ }\href@noop {} {\  (\bibinfo {year} {2023}{\natexlab{a}})},\
  \Eprint {http://arxiv.org/abs/2311.05503} {arXiv:2311.05503 [gr-qc]}
  \BibitemShut {NoStop}%
\bibitem [{\citenamefont {Bolokhov}(2023{\natexlab{b}})}]{Bolokhov:2023ruj}%
  \BibitemOpen
  \bibfield  {author} {\bibinfo {author} {\bibfnamefont {S.~V.}\ \bibnamefont
  {Bolokhov}},\ }\href {\doibase 10.20944/preprints202310.0517.v1} {\
  (\bibinfo {year} {2023}{\natexlab{b}}),\
  10.20944/preprints202310.0517.v1}\BibitemShut {NoStop}%
\bibitem [{\citenamefont {Konoplya}\ and\ \citenamefont
  {Fontana}(2008)}]{Konoplya:2007yy}%
  \BibitemOpen
  \bibfield  {author} {\bibinfo {author} {\bibfnamefont {R.~A.}\ \bibnamefont
  {Konoplya}}\ and\ \bibinfo {author} {\bibfnamefont {R.~D.~B.}\ \bibnamefont
  {Fontana}},\ }\href {\doibase 10.1016/j.physletb.2007.10.065} {\bibfield
  {journal} {\bibinfo  {journal} {Phys. Lett. B}\ }\textbf {\bibinfo {volume}
  {659}},\ \bibinfo {pages} {375} (\bibinfo {year} {2008})},\ \Eprint
  {http://arxiv.org/abs/0707.1156} {arXiv:0707.1156 [hep-th]} \BibitemShut
  {NoStop}%
\bibitem [{\citenamefont {Zinhailo}(2018)}]{Zinhailo:2018ska}%
  \BibitemOpen
  \bibfield  {author} {\bibinfo {author} {\bibfnamefont {A.~F.}\ \bibnamefont
  {Zinhailo}},\ }\href {\doibase 10.1140/epjc/s10052-018-6467-8} {\bibfield
  {journal} {\bibinfo  {journal} {Eur. Phys. J. C}\ }\textbf {\bibinfo {volume}
  {78}},\ \bibinfo {pages} {992} (\bibinfo {year} {2018})},\ \Eprint
  {http://arxiv.org/abs/1809.03913} {arXiv:1809.03913 [gr-qc]} \BibitemShut
  {NoStop}%
\bibitem [{\citenamefont {Kokkotas}\ \emph {et~al.}(2011)\citenamefont
  {Kokkotas}, \citenamefont {Konoplya},\ and\ \citenamefont
  {Zhidenko}}]{Kokkotas:2010zd}%
  \BibitemOpen
  \bibfield  {author} {\bibinfo {author} {\bibfnamefont {K.~D.}\ \bibnamefont
  {Kokkotas}}, \bibinfo {author} {\bibfnamefont {R.~A.}\ \bibnamefont
  {Konoplya}}, \ and\ \bibinfo {author} {\bibfnamefont {A.}~\bibnamefont
  {Zhidenko}},\ }\href {\doibase 10.1103/PhysRevD.83.024031} {\bibfield
  {journal} {\bibinfo  {journal} {Phys. Rev. D}\ }\textbf {\bibinfo {volume}
  {83}},\ \bibinfo {pages} {024031} (\bibinfo {year} {2011})},\ \Eprint
  {http://arxiv.org/abs/1011.1843} {arXiv:1011.1843 [gr-qc]} \BibitemShut
  {NoStop}%
\bibitem [{\citenamefont {Konoplya}\ and\ \citenamefont
  {Zinhailo}(2020)}]{Konoplya:2020bxa}%
  \BibitemOpen
  \bibfield  {author} {\bibinfo {author} {\bibfnamefont {R.~A.}\ \bibnamefont
  {Konoplya}}\ and\ \bibinfo {author} {\bibfnamefont {A.~F.}\ \bibnamefont
  {Zinhailo}},\ }\href {\doibase 10.1140/epjc/s10052-020-08639-8} {\bibfield
  {journal} {\bibinfo  {journal} {Eur. Phys. J. C}\ }\textbf {\bibinfo {volume}
  {80}},\ \bibinfo {pages} {1049} (\bibinfo {year} {2020})},\ \Eprint
  {http://arxiv.org/abs/2003.01188} {arXiv:2003.01188 [gr-qc]} \BibitemShut
  {NoStop}%
\bibitem [{\citenamefont {Konoplya}\ \emph
  {et~al.}(2019{\natexlab{a}})\citenamefont {Konoplya}, \citenamefont
  {Zinhailo},\ and\ \citenamefont {Stuchl\'\i{}k}}]{Konoplya:2019hml}%
  \BibitemOpen
  \bibfield  {author} {\bibinfo {author} {\bibfnamefont {R.~A.}\ \bibnamefont
  {Konoplya}}, \bibinfo {author} {\bibfnamefont {A.~F.}\ \bibnamefont
  {Zinhailo}}, \ and\ \bibinfo {author} {\bibfnamefont {Z.}~\bibnamefont
  {Stuchl\'\i{}k}},\ }\href {\doibase 10.1103/PhysRevD.99.124042} {\bibfield
  {journal} {\bibinfo  {journal} {Phys. Rev. D}\ }\textbf {\bibinfo {volume}
  {99}},\ \bibinfo {pages} {124042} (\bibinfo {year} {2019}{\natexlab{a}})},\
  \Eprint {http://arxiv.org/abs/1903.03483} {arXiv:1903.03483 [gr-qc]}
  \BibitemShut {NoStop}%
\bibitem [{\citenamefont {Chen}\ \emph {et~al.}(2023)\citenamefont {Chen},
  \citenamefont {Pan},\ and\ \citenamefont {Jing}}]{Chen:2023cjd}%
  \BibitemOpen
  \bibfield  {author} {\bibinfo {author} {\bibfnamefont {C.}~\bibnamefont
  {Chen}}, \bibinfo {author} {\bibfnamefont {Q.}~\bibnamefont {Pan}}, \ and\
  \bibinfo {author} {\bibfnamefont {J.}~\bibnamefont {Jing}},\ }\href {\doibase
  10.1016/j.physletb.2023.138186} {\bibfield  {journal} {\bibinfo  {journal}
  {Phys. Lett. B}\ }\textbf {\bibinfo {volume} {846}},\ \bibinfo {pages}
  {138186} (\bibinfo {year} {2023})},\ \Eprint
  {http://arxiv.org/abs/2302.05861} {arXiv:2302.05861 [gr-qc]} \BibitemShut
  {NoStop}%
\bibitem [{\citenamefont {Panotopoulos}(2018)}]{Panotopoulos:2018can}%
  \BibitemOpen
  \bibfield  {author} {\bibinfo {author} {\bibfnamefont {G.}~\bibnamefont
  {Panotopoulos}},\ }\href {\doibase 10.1007/s10714-018-2381-5} {\bibfield
  {journal} {\bibinfo  {journal} {Gen. Rel. Grav.}\ }\textbf {\bibinfo {volume}
  {50}},\ \bibinfo {pages} {59} (\bibinfo {year} {2018})},\ \Eprint
  {http://arxiv.org/abs/1805.04743} {arXiv:1805.04743 [hep-th]} \BibitemShut
  {NoStop}%
\bibitem [{\citenamefont {Huang}\ \emph {et~al.}(2018)\citenamefont {Huang},
  \citenamefont {Chen},\ and\ \citenamefont {Wang}}]{Huang:2018vlq}%
  \BibitemOpen
  \bibfield  {author} {\bibinfo {author} {\bibfnamefont {L.}~\bibnamefont
  {Huang}}, \bibinfo {author} {\bibfnamefont {J.}~\bibnamefont {Chen}}, \ and\
  \bibinfo {author} {\bibfnamefont {Y.}~\bibnamefont {Wang}},\ }\href {\doibase
  10.1140/epjc/s10052-018-5779-z} {\bibfield  {journal} {\bibinfo  {journal}
  {Eur. Phys. J. C}\ }\textbf {\bibinfo {volume} {78}},\ \bibinfo {pages} {299}
  (\bibinfo {year} {2018})}\BibitemShut {NoStop}%
\bibitem [{\citenamefont {Gupta}\ \emph {et~al.}(2017)\citenamefont {Gupta},
  \citenamefont {Juri\'c},\ and\ \citenamefont {Samsarov}}]{Gupta:2017lwk}%
  \BibitemOpen
  \bibfield  {author} {\bibinfo {author} {\bibfnamefont {K.~S.}\ \bibnamefont
  {Gupta}}, \bibinfo {author} {\bibfnamefont {T.}~\bibnamefont {Juri\'c}}, \
  and\ \bibinfo {author} {\bibfnamefont {A.}~\bibnamefont {Samsarov}},\ }\href
  {\doibase 10.1007/JHEP06(2017)107} {\bibfield  {journal} {\bibinfo  {journal}
  {JHEP}\ }\textbf {\bibinfo {volume} {06}},\ \bibinfo {pages} {107} (\bibinfo
  {year} {2017})},\ \Eprint {http://arxiv.org/abs/1703.00514} {arXiv:1703.00514
  [hep-th]} \BibitemShut {NoStop}%
\bibitem [{\citenamefont {Prasia}\ and\ \citenamefont
  {Kuriakose}(2017)}]{Prasia:2016esx}%
  \BibitemOpen
  \bibfield  {author} {\bibinfo {author} {\bibfnamefont {P.}~\bibnamefont
  {Prasia}}\ and\ \bibinfo {author} {\bibfnamefont {V.~C.}\ \bibnamefont
  {Kuriakose}},\ }\href {\doibase 10.1140/epjc/s10052-016-4591-x} {\bibfield
  {journal} {\bibinfo  {journal} {Eur. Phys. J. C}\ }\textbf {\bibinfo {volume}
  {77}},\ \bibinfo {pages} {27} (\bibinfo {year} {2017})},\ \Eprint
  {http://arxiv.org/abs/1608.05299} {arXiv:1608.05299 [gr-qc]} \BibitemShut
  {NoStop}%
\bibitem [{\citenamefont {Becar}\ \emph {et~al.}(2014)\citenamefont {Becar},
  \citenamefont {Gonzalez},\ and\ \citenamefont {Vasquez}}]{Becar:2013qba}%
  \BibitemOpen
  \bibfield  {author} {\bibinfo {author} {\bibfnamefont {R.}~\bibnamefont
  {Becar}}, \bibinfo {author} {\bibfnamefont {P.~A.}\ \bibnamefont {Gonzalez}},
  \ and\ \bibinfo {author} {\bibfnamefont {Y.}~\bibnamefont {Vasquez}},\ }\href
  {\doibase 10.1103/PhysRevD.89.023001} {\bibfield  {journal} {\bibinfo
  {journal} {Phys. Rev. D}\ }\textbf {\bibinfo {volume} {89}},\ \bibinfo
  {pages} {023001} (\bibinfo {year} {2014})},\ \Eprint
  {http://arxiv.org/abs/1306.5974} {arXiv:1306.5974 [gr-qc]} \BibitemShut
  {NoStop}%
\bibitem [{\citenamefont {Kim}\ \emph {et~al.}(2012)\citenamefont {Kim},
  \citenamefont {Myung},\ and\ \citenamefont {Park}}]{Kim:2012pt}%
  \BibitemOpen
  \bibfield  {author} {\bibinfo {author} {\bibfnamefont {Y.-W.}\ \bibnamefont
  {Kim}}, \bibinfo {author} {\bibfnamefont {Y.~S.}\ \bibnamefont {Myung}}, \
  and\ \bibinfo {author} {\bibfnamefont {Y.-J.}\ \bibnamefont {Park}},\ }\href
  {\doibase 10.1103/PhysRevD.85.124018} {\bibfield  {journal} {\bibinfo
  {journal} {Phys. Rev. D}\ }\textbf {\bibinfo {volume} {85}},\ \bibinfo
  {pages} {124018} (\bibinfo {year} {2012})},\ \Eprint
  {http://arxiv.org/abs/1204.3706} {arXiv:1204.3706 [hep-th]} \BibitemShut
  {NoStop}%
\bibitem [{\citenamefont {Myung}\ \emph {et~al.}(2012)\citenamefont {Myung},
  \citenamefont {Kim},\ and\ \citenamefont {Park}}]{Myung:2012sh}%
  \BibitemOpen
  \bibfield  {author} {\bibinfo {author} {\bibfnamefont {Y.~S.}\ \bibnamefont
  {Myung}}, \bibinfo {author} {\bibfnamefont {Y.-W.}\ \bibnamefont {Kim}}, \
  and\ \bibinfo {author} {\bibfnamefont {Y.-J.}\ \bibnamefont {Park}},\ }\href
  {\doibase 10.1103/PhysRevD.85.084007} {\bibfield  {journal} {\bibinfo
  {journal} {Phys. Rev. D}\ }\textbf {\bibinfo {volume} {85}},\ \bibinfo
  {pages} {084007} (\bibinfo {year} {2012})},\ \Eprint
  {http://arxiv.org/abs/1201.3964} {arXiv:1201.3964 [hep-th]} \BibitemShut
  {NoStop}%
\bibitem [{\citenamefont {Schutz}\ and\ \citenamefont {Will}(1985)}]{wkb1}%
  \BibitemOpen
  \bibfield  {author} {\bibinfo {author} {\bibfnamefont {B.~F.}\ \bibnamefont
  {Schutz}}\ and\ \bibinfo {author} {\bibfnamefont {C.~M.}\ \bibnamefont
  {Will}},\ }\href {\doibase 10.1086/184453} {\bibfield  {journal} {\bibinfo
  {journal} {Astrophys. J.}\ }\textbf {\bibinfo {volume} {291}},\ \bibinfo
  {pages} {L33} (\bibinfo {year} {1985})}\BibitemShut {NoStop}%
\bibitem [{\citenamefont {Iyer}\ and\ \citenamefont {Will}(1987)}]{wkb2}%
  \BibitemOpen
  \bibfield  {author} {\bibinfo {author} {\bibfnamefont {S.}~\bibnamefont
  {Iyer}}\ and\ \bibinfo {author} {\bibfnamefont {C.~M.}\ \bibnamefont
  {Will}},\ }\href {\doibase 10.1103/PhysRevD.35.3621} {\bibfield  {journal}
  {\bibinfo  {journal} {Phys. Rev.}\ }\textbf {\bibinfo {volume} {D35}},\
  \bibinfo {pages} {3621} (\bibinfo {year} {1987})}\BibitemShut {NoStop}%
\bibitem [{\citenamefont {Konoplya}(2003)}]{Konoplya:2003ii}%
  \BibitemOpen
  \bibfield  {author} {\bibinfo {author} {\bibfnamefont {R.~A.}\ \bibnamefont
  {Konoplya}},\ }\href {\doibase 10.1103/PhysRevD.68.024018} {\bibfield
  {journal} {\bibinfo  {journal} {Phys. Rev. D}\ }\textbf {\bibinfo {volume}
  {68}},\ \bibinfo {pages} {024018} (\bibinfo {year} {2003})},\ \Eprint
  {http://arxiv.org/abs/gr-qc/0303052} {arXiv:gr-qc/0303052} \BibitemShut
  {NoStop}%
\bibitem [{\citenamefont {Gundlach}\ \emph {et~al.}(1994)\citenamefont
  {Gundlach}, \citenamefont {Price},\ and\ \citenamefont
  {Pullin}}]{Gundlach:1993tp}%
  \BibitemOpen
  \bibfield  {author} {\bibinfo {author} {\bibfnamefont {C.}~\bibnamefont
  {Gundlach}}, \bibinfo {author} {\bibfnamefont {R.~H.}\ \bibnamefont {Price}},
  \ and\ \bibinfo {author} {\bibfnamefont {J.}~\bibnamefont {Pullin}},\ }\href
  {\doibase 10.1103/PhysRevD.49.883} {\bibfield  {journal} {\bibinfo  {journal}
  {Phys. Rev. D}\ }\textbf {\bibinfo {volume} {49}},\ \bibinfo {pages} {883}
  (\bibinfo {year} {1994})},\ \Eprint {http://arxiv.org/abs/gr-qc/9307009}
  {arXiv:gr-qc/9307009} \BibitemShut {NoStop}%
\bibitem [{\citenamefont {Churilova}\ and\ \citenamefont
  {Stuchlik}(2020)}]{Churilova:2019cyt}%
  \BibitemOpen
  \bibfield  {author} {\bibinfo {author} {\bibfnamefont {M.~S.}\ \bibnamefont
  {Churilova}}\ and\ \bibinfo {author} {\bibfnamefont {Z.}~\bibnamefont
  {Stuchlik}},\ }\href {\doibase 10.1088/1361-6382/ab7717} {\bibfield
  {journal} {\bibinfo  {journal} {Class. Quant. Grav.}\ }\textbf {\bibinfo
  {volume} {37}},\ \bibinfo {pages} {075014} (\bibinfo {year} {2020})},\
  \Eprint {http://arxiv.org/abs/1911.11823} {arXiv:1911.11823 [gr-qc]}
  \BibitemShut {NoStop}%
\bibitem [{\citenamefont {Bronnikov}\ and\ \citenamefont
  {Konoplya}(2020)}]{Bronnikov:2019sbx}%
  \BibitemOpen
  \bibfield  {author} {\bibinfo {author} {\bibfnamefont {K.~A.}\ \bibnamefont
  {Bronnikov}}\ and\ \bibinfo {author} {\bibfnamefont {R.~A.}\ \bibnamefont
  {Konoplya}},\ }\href {\doibase 10.1103/PhysRevD.101.064004} {\bibfield
  {journal} {\bibinfo  {journal} {Phys. Rev. D}\ }\textbf {\bibinfo {volume}
  {101}},\ \bibinfo {pages} {064004} (\bibinfo {year} {2020})},\ \Eprint
  {http://arxiv.org/abs/1912.05315} {arXiv:1912.05315 [gr-qc]} \BibitemShut
  {NoStop}%
\bibitem [{\citenamefont {Wang}\ \emph {et~al.}(2004)\citenamefont {Wang},
  \citenamefont {Lin},\ and\ \citenamefont {Molina}}]{Wang:2004bv}%
  \BibitemOpen
  \bibfield  {author} {\bibinfo {author} {\bibfnamefont {B.}~\bibnamefont
  {Wang}}, \bibinfo {author} {\bibfnamefont {C.-Y.}\ \bibnamefont {Lin}}, \
  and\ \bibinfo {author} {\bibfnamefont {C.}~\bibnamefont {Molina}},\ }\href
  {\doibase 10.1103/PhysRevD.70.064025} {\bibfield  {journal} {\bibinfo
  {journal} {Phys. Rev. D}\ }\textbf {\bibinfo {volume} {70}},\ \bibinfo
  {pages} {064025} (\bibinfo {year} {2004})},\ \Eprint
  {http://arxiv.org/abs/hep-th/0407024} {arXiv:hep-th/0407024} \BibitemShut
  {NoStop}%
\bibitem [{\citenamefont {Konoplya}\ \emph
  {et~al.}(2019{\natexlab{b}})\citenamefont {Konoplya}, \citenamefont
  {Zhidenko},\ and\ \citenamefont {Zinhailo}}]{Konoplya:2019hlu}%
  \BibitemOpen
  \bibfield  {author} {\bibinfo {author} {\bibfnamefont {R.~A.}\ \bibnamefont
  {Konoplya}}, \bibinfo {author} {\bibfnamefont {A.}~\bibnamefont {Zhidenko}},
  \ and\ \bibinfo {author} {\bibfnamefont {A.~F.}\ \bibnamefont {Zinhailo}},\
  }\href {\doibase 10.1088/1361-6382/ab2e25} {\bibfield  {journal} {\bibinfo
  {journal} {Class. Quant. Grav.}\ }\textbf {\bibinfo {volume} {36}},\ \bibinfo
  {pages} {155002} (\bibinfo {year} {2019}{\natexlab{b}})},\ \Eprint
  {http://arxiv.org/abs/1904.10333} {arXiv:1904.10333 [gr-qc]} \BibitemShut
  {NoStop}%
\bibitem [{\citenamefont {Matyjasek}\ and\ \citenamefont
  {Opala}(2017)}]{Matyjasek:2017psv}%
  \BibitemOpen
  \bibfield  {author} {\bibinfo {author} {\bibfnamefont {J.}~\bibnamefont
  {Matyjasek}}\ and\ \bibinfo {author} {\bibfnamefont {M.}~\bibnamefont
  {Opala}},\ }\href {\doibase 10.1103/PhysRevD.96.024011} {\bibfield  {journal}
  {\bibinfo  {journal} {Phys. Rev. D}\ }\textbf {\bibinfo {volume} {96}},\
  \bibinfo {pages} {024011} (\bibinfo {year} {2017})},\ \Eprint
  {http://arxiv.org/abs/1704.00361} {arXiv:1704.00361 [gr-qc]} \BibitemShut
  {NoStop}%
\bibitem [{\citenamefont {Bolokhov}(2023{\natexlab{c}})}]{Bolokhov:2023dxq}%
  \BibitemOpen
  \bibfield  {author} {\bibinfo {author} {\bibfnamefont {S.~V.}\ \bibnamefont
  {Bolokhov}},\ }\href@noop {} {\  (\bibinfo {year} {2023}{\natexlab{c}})},\
  \Eprint {http://arxiv.org/abs/2310.12326} {arXiv:2310.12326 [gr-qc]}
  \BibitemShut {NoStop}%
\bibitem [{\citenamefont {Fortuna}\ and\ \citenamefont
  {Vega}(2020)}]{Fortuna:2020obg}%
  \BibitemOpen
  \bibfield  {author} {\bibinfo {author} {\bibfnamefont {S.}~\bibnamefont
  {Fortuna}}\ and\ \bibinfo {author} {\bibfnamefont {I.}~\bibnamefont {Vega}},\
  }\href@noop {} {\  (\bibinfo {year} {2020})},\ \Eprint
  {http://arxiv.org/abs/2003.06232} {arXiv:2003.06232 [gr-qc]} \BibitemShut
  {NoStop}%
\bibitem [{\citenamefont {Konoplya}\ \emph {et~al.}(2023)\citenamefont
  {Konoplya}, \citenamefont {Stuchlik}, \citenamefont {Zhidenko},\ and\
  \citenamefont {Zinhailo}}]{Konoplya:2023aph}%
  \BibitemOpen
  \bibfield  {author} {\bibinfo {author} {\bibfnamefont {R.~A.}\ \bibnamefont
  {Konoplya}}, \bibinfo {author} {\bibfnamefont {Z.}~\bibnamefont {Stuchlik}},
  \bibinfo {author} {\bibfnamefont {A.}~\bibnamefont {Zhidenko}}, \ and\
  \bibinfo {author} {\bibfnamefont {A.~F.}\ \bibnamefont {Zinhailo}},\ }\href
  {\doibase 10.1103/PhysRevD.107.104050} {\bibfield  {journal} {\bibinfo
  {journal} {Phys. Rev. D}\ }\textbf {\bibinfo {volume} {107}},\ \bibinfo
  {pages} {104050} (\bibinfo {year} {2023})},\ \Eprint
  {http://arxiv.org/abs/2303.01987} {arXiv:2303.01987 [gr-qc]} \BibitemShut
  {NoStop}%
\bibitem [{\citenamefont {Konoplya}\ and\ \citenamefont
  {Zhidenko}(2023)}]{Konoplya:2022zav}%
  \BibitemOpen
  \bibfield  {author} {\bibinfo {author} {\bibfnamefont {R.~A.}\ \bibnamefont
  {Konoplya}}\ and\ \bibinfo {author} {\bibfnamefont {A.}~\bibnamefont
  {Zhidenko}},\ }\href {\doibase 10.1103/PhysRevD.107.044009} {\bibfield
  {journal} {\bibinfo  {journal} {Phys. Rev. D}\ }\textbf {\bibinfo {volume}
  {107}},\ \bibinfo {pages} {044009} (\bibinfo {year} {2023})},\ \Eprint
  {http://arxiv.org/abs/2211.02997} {arXiv:2211.02997 [gr-qc]} \BibitemShut
  {NoStop}%
\bibitem [{\citenamefont {Konoplya}\ \emph {et~al.}(2008)\citenamefont
  {Konoplya}, \citenamefont {Murata}, \citenamefont {Soda},\ and\ \citenamefont
  {Zhidenko}}]{Konoplya:2008yy}%
  \BibitemOpen
  \bibfield  {author} {\bibinfo {author} {\bibfnamefont {R.~A.}\ \bibnamefont
  {Konoplya}}, \bibinfo {author} {\bibfnamefont {K.}~\bibnamefont {Murata}},
  \bibinfo {author} {\bibfnamefont {J.}~\bibnamefont {Soda}}, \ and\ \bibinfo
  {author} {\bibfnamefont {A.}~\bibnamefont {Zhidenko}},\ }\href {\doibase
  10.1103/PhysRevD.78.084012} {\bibfield  {journal} {\bibinfo  {journal} {Phys.
  Rev. D}\ }\textbf {\bibinfo {volume} {78}},\ \bibinfo {pages} {084012}
  (\bibinfo {year} {2008})},\ \Eprint {http://arxiv.org/abs/0807.1897}
  {arXiv:0807.1897 [hep-th]} \BibitemShut {NoStop}%
\bibitem [{\citenamefont {Zhidenko}(2006)}]{Zhidenko:2006rs}%
  \BibitemOpen
  \bibfield  {author} {\bibinfo {author} {\bibfnamefont {A.}~\bibnamefont
  {Zhidenko}},\ }\href {\doibase 10.1103/PhysRevD.74.064017} {\bibfield
  {journal} {\bibinfo  {journal} {Phys. Rev. D}\ }\textbf {\bibinfo {volume}
  {74}},\ \bibinfo {pages} {064017} (\bibinfo {year} {2006})},\ \Eprint
  {http://arxiv.org/abs/gr-qc/0607133} {arXiv:gr-qc/0607133} \BibitemShut
  {NoStop}%
\bibitem [{\citenamefont {Leaver}(1985)}]{Leaver:1985ax}%
  \BibitemOpen
  \bibfield  {author} {\bibinfo {author} {\bibfnamefont {E.~W.}\ \bibnamefont
  {Leaver}},\ }\href {\doibase 10.1098/rspa.1985.0119} {\bibfield  {journal}
  {\bibinfo  {journal} {Proc. Roy. Soc. Lond. A}\ }\textbf {\bibinfo {volume}
  {402}},\ \bibinfo {pages} {285} (\bibinfo {year} {1985})}\BibitemShut
  {NoStop}%
\bibitem [{\citenamefont {Price}(1972)}]{Price:1972pw}%
  \BibitemOpen
  \bibfield  {author} {\bibinfo {author} {\bibfnamefont {R.~H.}\ \bibnamefont
  {Price}},\ }\href {\doibase 10.1103/PhysRevD.5.2439} {\bibfield  {journal}
  {\bibinfo  {journal} {Phys. Rev. D}\ }\textbf {\bibinfo {volume} {5}},\
  \bibinfo {pages} {2439} (\bibinfo {year} {1972})}\BibitemShut {NoStop}%
\bibitem [{\citenamefont {Koyama}\ and\ \citenamefont
  {Tomimatsu}(2001)}]{Koyama:2000hj}%
  \BibitemOpen
  \bibfield  {author} {\bibinfo {author} {\bibfnamefont {H.}~\bibnamefont
  {Koyama}}\ and\ \bibinfo {author} {\bibfnamefont {A.}~\bibnamefont
  {Tomimatsu}},\ }\href {\doibase 10.1103/PhysRevD.63.064032} {\bibfield
  {journal} {\bibinfo  {journal} {Phys. Rev. D}\ }\textbf {\bibinfo {volume}
  {63}},\ \bibinfo {pages} {064032} (\bibinfo {year} {2001})},\ \Eprint
  {http://arxiv.org/abs/gr-qc/0012022} {arXiv:gr-qc/0012022} \BibitemShut
  {NoStop}%
\bibitem [{\citenamefont {Koyama}\ and\ \citenamefont
  {Tomimatsu}(2002)}]{Koyama:2001qw}%
  \BibitemOpen
  \bibfield  {author} {\bibinfo {author} {\bibfnamefont {H.}~\bibnamefont
  {Koyama}}\ and\ \bibinfo {author} {\bibfnamefont {A.}~\bibnamefont
  {Tomimatsu}},\ }\href {\doibase 10.1103/PhysRevD.65.084031} {\bibfield
  {journal} {\bibinfo  {journal} {Phys. Rev. D}\ }\textbf {\bibinfo {volume}
  {65}},\ \bibinfo {pages} {084031} (\bibinfo {year} {2002})},\ \Eprint
  {http://arxiv.org/abs/gr-qc/0112075} {arXiv:gr-qc/0112075} \BibitemShut
  {NoStop}%
\bibitem [{\citenamefont {Moderski}\ and\ \citenamefont
  {Rogatko}(2001)}]{Moderski:2001tk}%
  \BibitemOpen
  \bibfield  {author} {\bibinfo {author} {\bibfnamefont {R.}~\bibnamefont
  {Moderski}}\ and\ \bibinfo {author} {\bibfnamefont {M.}~\bibnamefont
  {Rogatko}},\ }\href {\doibase 10.1103/PhysRevD.64.044024} {\bibfield
  {journal} {\bibinfo  {journal} {Phys. Rev. D}\ }\textbf {\bibinfo {volume}
  {64}},\ \bibinfo {pages} {044024} (\bibinfo {year} {2001})},\ \Eprint
  {http://arxiv.org/abs/gr-qc/0105056} {arXiv:gr-qc/0105056} \BibitemShut
  {NoStop}%
\bibitem [{\citenamefont {Konoplya}\ \emph {et~al.}(2007)\citenamefont
  {Konoplya}, \citenamefont {Zhidenko},\ and\ \citenamefont
  {Molina}}]{Konoplya:2006gq}%
  \BibitemOpen
  \bibfield  {author} {\bibinfo {author} {\bibfnamefont {R.~A.}\ \bibnamefont
  {Konoplya}}, \bibinfo {author} {\bibfnamefont {A.}~\bibnamefont {Zhidenko}},
  \ and\ \bibinfo {author} {\bibfnamefont {C.}~\bibnamefont {Molina}},\ }\href
  {\doibase 10.1103/PhysRevD.75.084004} {\bibfield  {journal} {\bibinfo
  {journal} {Phys. Rev. D}\ }\textbf {\bibinfo {volume} {75}},\ \bibinfo
  {pages} {084004} (\bibinfo {year} {2007})},\ \Eprint
  {http://arxiv.org/abs/gr-qc/0602047} {arXiv:gr-qc/0602047} \BibitemShut
  {NoStop}%
\bibitem [{\citenamefont {Jing}(2005)}]{Jing:2004zb}%
  \BibitemOpen
  \bibfield  {author} {\bibinfo {author} {\bibfnamefont {J.}~\bibnamefont
  {Jing}},\ }\href {\doibase 10.1103/PhysRevD.72.027501} {\bibfield  {journal}
  {\bibinfo  {journal} {Phys. Rev. D}\ }\textbf {\bibinfo {volume} {72}},\
  \bibinfo {pages} {027501} (\bibinfo {year} {2005})},\ \Eprint
  {http://arxiv.org/abs/gr-qc/0408090} {arXiv:gr-qc/0408090} \BibitemShut
  {NoStop}%
\bibitem [{\citenamefont {Seahra}\ \emph {et~al.}(2005)\citenamefont {Seahra},
  \citenamefont {Clarkson},\ and\ \citenamefont {Maartens}}]{Seahra:2004fg}%
  \BibitemOpen
  \bibfield  {author} {\bibinfo {author} {\bibfnamefont {S.~S.}\ \bibnamefont
  {Seahra}}, \bibinfo {author} {\bibfnamefont {C.}~\bibnamefont {Clarkson}}, \
  and\ \bibinfo {author} {\bibfnamefont {R.}~\bibnamefont {Maartens}},\ }\href
  {\doibase 10.1103/PhysRevLett.94.121302} {\bibfield  {journal} {\bibinfo
  {journal} {Phys. Rev. Lett.}\ }\textbf {\bibinfo {volume} {94}},\ \bibinfo
  {pages} {121302} (\bibinfo {year} {2005})},\ \Eprint
  {http://arxiv.org/abs/gr-qc/0408032} {arXiv:gr-qc/0408032} \BibitemShut
  {NoStop}%
\bibitem [{\citenamefont {Zhidenko}(2004)}]{Zhidenko:2003wq}%
  \BibitemOpen
  \bibfield  {author} {\bibinfo {author} {\bibfnamefont {A.}~\bibnamefont
  {Zhidenko}},\ }\href {\doibase 10.1088/0264-9381/21/1/019} {\bibfield
  {journal} {\bibinfo  {journal} {Class. Quant. Grav.}\ }\textbf {\bibinfo
  {volume} {21}},\ \bibinfo {pages} {273} (\bibinfo {year} {2004})},\ \Eprint
  {http://arxiv.org/abs/gr-qc/0307012} {arXiv:gr-qc/0307012} \BibitemShut
  {NoStop}%
\bibitem [{\citenamefont {Konoplya}\ and\ \citenamefont
  {Zhidenko}(2004)}]{Konoplya:2004uk}%
  \BibitemOpen
  \bibfield  {author} {\bibinfo {author} {\bibfnamefont {R.~A.}\ \bibnamefont
  {Konoplya}}\ and\ \bibinfo {author} {\bibfnamefont {A.}~\bibnamefont
  {Zhidenko}},\ }\href {\doibase 10.1088/1126-6708/2004/06/037} {\bibfield
  {journal} {\bibinfo  {journal} {JHEP}\ }\textbf {\bibinfo {volume} {06}},\
  \bibinfo {pages} {037} (\bibinfo {year} {2004})},\ \Eprint
  {http://arxiv.org/abs/hep-th/0402080} {arXiv:hep-th/0402080} \BibitemShut
  {NoStop}%
\bibitem [{\citenamefont {Giammatteo}\ and\ \citenamefont
  {Moss}(2005)}]{Giammatteo:2005vu}%
  \BibitemOpen
  \bibfield  {author} {\bibinfo {author} {\bibfnamefont {M.}~\bibnamefont
  {Giammatteo}}\ and\ \bibinfo {author} {\bibfnamefont {I.~G.}\ \bibnamefont
  {Moss}},\ }\href {\doibase 10.1088/0264-9381/22/9/021} {\bibfield  {journal}
  {\bibinfo  {journal} {Class. Quant. Grav.}\ }\textbf {\bibinfo {volume}
  {22}},\ \bibinfo {pages} {1803} (\bibinfo {year} {2005})},\ \Eprint
  {http://arxiv.org/abs/gr-qc/0502046} {arXiv:gr-qc/0502046} \BibitemShut
  {NoStop}%
\bibitem [{\citenamefont {Molina}\ \emph {et~al.}(2004)\citenamefont {Molina},
  \citenamefont {Giugno}, \citenamefont {Abdalla},\ and\ \citenamefont
  {Saa}}]{Molina:2003dc}%
  \BibitemOpen
  \bibfield  {author} {\bibinfo {author} {\bibfnamefont {C.}~\bibnamefont
  {Molina}}, \bibinfo {author} {\bibfnamefont {D.}~\bibnamefont {Giugno}},
  \bibinfo {author} {\bibfnamefont {E.}~\bibnamefont {Abdalla}}, \ and\
  \bibinfo {author} {\bibfnamefont {A.}~\bibnamefont {Saa}},\ }\href {\doibase
  10.1103/PhysRevD.69.104013} {\bibfield  {journal} {\bibinfo  {journal} {Phys.
  Rev. D}\ }\textbf {\bibinfo {volume} {69}},\ \bibinfo {pages} {104013}
  (\bibinfo {year} {2004})},\ \Eprint {http://arxiv.org/abs/gr-qc/0309079}
  {arXiv:gr-qc/0309079} \BibitemShut {NoStop}%
\bibitem [{\citenamefont {Brady}\ \emph {et~al.}(1999)\citenamefont {Brady},
  \citenamefont {Chambers}, \citenamefont {Laarakkers},\ and\ \citenamefont
  {Poisson}}]{Brady:1999wd}%
  \BibitemOpen
  \bibfield  {author} {\bibinfo {author} {\bibfnamefont {P.~R.}\ \bibnamefont
  {Brady}}, \bibinfo {author} {\bibfnamefont {C.~M.}\ \bibnamefont {Chambers}},
  \bibinfo {author} {\bibfnamefont {W.~G.}\ \bibnamefont {Laarakkers}}, \ and\
  \bibinfo {author} {\bibfnamefont {E.}~\bibnamefont {Poisson}},\ }\href
  {\doibase 10.1103/PhysRevD.60.064003} {\bibfield  {journal} {\bibinfo
  {journal} {Phys. Rev. D}\ }\textbf {\bibinfo {volume} {60}},\ \bibinfo
  {pages} {064003} (\bibinfo {year} {1999})},\ \Eprint
  {http://arxiv.org/abs/gr-qc/9902010} {arXiv:gr-qc/9902010} \BibitemShut
  {NoStop}%
\bibitem [{\citenamefont {Konoplya}\ and\ \citenamefont
  {Zhidenko}(2022)}]{Konoplya:2022xid}%
  \BibitemOpen
  \bibfield  {author} {\bibinfo {author} {\bibfnamefont {R.~A.}\ \bibnamefont
  {Konoplya}}\ and\ \bibinfo {author} {\bibfnamefont {A.}~\bibnamefont
  {Zhidenko}},\ }\href {\doibase 10.1103/PhysRevD.106.124004} {\bibfield
  {journal} {\bibinfo  {journal} {Phys. Rev. D}\ }\textbf {\bibinfo {volume}
  {106}},\ \bibinfo {pages} {124004} (\bibinfo {year} {2022})},\ \Eprint
  {http://arxiv.org/abs/2209.12058} {arXiv:2209.12058 [gr-qc]} \BibitemShut
  {NoStop}%
\bibitem [{\citenamefont {Konoplya}\ and\ \citenamefont
  {Zhidenko}(2005)}]{Konoplya:2004wg}%
  \BibitemOpen
  \bibfield  {author} {\bibinfo {author} {\bibfnamefont {R.~A.}\ \bibnamefont
  {Konoplya}}\ and\ \bibinfo {author} {\bibfnamefont {A.~V.}\ \bibnamefont
  {Zhidenko}},\ }\href {\doibase 10.1016/j.physletb.2005.01.078} {\bibfield
  {journal} {\bibinfo  {journal} {Phys. Lett. B}\ }\textbf {\bibinfo {volume}
  {609}},\ \bibinfo {pages} {377} (\bibinfo {year} {2005})},\ \Eprint
  {http://arxiv.org/abs/gr-qc/0411059} {arXiv:gr-qc/0411059} \BibitemShut
  {NoStop}%
\end{thebibliography}%

\end{document}